\documentclass[epj]{svjour}
\usepackage{graphicx}

\def\mb#1{\setbox0=\hbox{$#1$}\kern-.025em\copy0\kern-\wd0
\kern-0.05em\copy0\kern-\wd0\kern-.025em\raise.0233em\box0}

\begin{document}
   \title{Phase separation of bacterial colonies in a limit of high 
degradation. Analogy with Jupiter's great red spot. }

 \author{P.H. Chavanis}

\institute{ Laboratoire de Physique Th\'eorique, Universit\'e Paul
Sabatier, 118 route de Narbonne 31062 Toulouse, France\\
\email{chavanis@irsamc.ups-tlse.fr}}

\titlerunning{Phase separation of bacterial colonies in a limit of high 
degradation.}

   \date{To be included later }

   \abstract{We discuss the structure of the equilibrium states of a
   regularized Keller-Segel model describing the chemotaxis of
   bacterial populations. We consider the limit of high degradation of
   the secreted chemical where analytical results can be
   obtained. Below a critical effective temperature, the system
   experiences a second order phase transition from a homogeneous
   phase to an inhomogeneous phase formed by two domains with uniform
   concentration separated by a thin interface (domain wall). We study
   the properties of the interface and determine the bifurcation
   between a circular shape (spot) and a stripe as a function of the control
   parameters. We show the analogy with the structure of Jupiter's
   Great red spot which also consists of two phases with uniform
   potential vorticity separated by a thin annular jet.  \PACS{
   {05.20.-y}{Classical statistical mechanics} \and
   {05.45.-a}{Nonlinear dynamics and nonlinear dynamical systems }} }

   \maketitle
%

\section{Introduction}
\label{sec_introduction}

The name chemotaxis refers to the motion of organisms induced by
chemical signals \cite{murray}. In some cases, the biological
organisms secrete a substance (pheromone, smell, food, ...) that has
an attractive effect on the organisms themselves. Therefore, in
addition to their diffusive motion, they move systematically along the
gradient of concentration of the chemical they secrete (chemotactic
flux). When attraction prevails over diffusion, the chemotaxis can
trigger a self-accelerating process until a point at which aggregation
takes place. This is the case for the slime mold {\it Dictyostelium
Discoideum} and for the bacteria {\it Escherichia coli}. A model of
slime mold aggregation has been introduced by Patlak \cite{patlak} and
Keller \& Segel \cite{ks} in the form of two coupled differential
equations. A simplified version of this model has been extensively
studied in the case where the degradation of the secreted chemical can
be neglected. In that case, the Keller-Segel equations become
isomorphic to the Smoluchowski-Poisson system describing
self-gravitating Brownian particles \cite{crrs}.  An analytical study
of this system of equations has been performed by Chavanis \& Sire
\cite{crs,sc,lang,post,tcoll,banach,sopik,virial} in a series of
papers. In particular, self-similar or quasi self-similar solutions
describing the chemotactic collapse have been obtained and the
formation of a Dirac peak has been found in the post-collapse regime
in $d>2$ and in the collapse regime in $d=2$. In parallel, a vast
number of rigorous results concerning the existence and unicity of
solutions of the Keller-Segel model and the conditions of blow-up have
been obtained by applied mathematicians (we refer to
\cite{horstmann} for a connection to the mathematical literature).

In this paper, we consider novel aspects of the Keller-Segel model. We
first introduce a regularized model that prevents finite time-blow up
and the formation of (unphysical) singularities like infinite density
profiles and Dirac peaks. In this model, the local density of cells is
bounded by a maximum value $\rho({\bf r})\le\sigma_{0}$ which takes
into account finite size effects and filling factors. Therefore, the
Dirac peaks (singularities) are replaced by smooth density profiles
(clumps).  With this regularization, there exists steady solutions
(similar to the Fermi-Dirac distribution) for any value of the control
parameter while the usual Keller-Segel model blows up above a critical
mass $M>M_{c}$ (in dimension $d\ge 2$).  In addition, we consider the
limit of high degradation of the chemical. In this limit, we show
that, for sufficiently small (effective) temperatures $T<T_{c}$, the
system undergoes a second order phase transition from a homogeneous
phase to an inhomogeneous phase. The bacteria organize in two domains
with uniform density $\rho_{\pm}$ separated by a thin interface. The
resulting structure is similar to a ``domain wall'' in phase ordering
kinetics \cite{bray}. We study in detail the structure of the
interface (profile, width, surface tension,...)  and determine the
conditions for the bifurcation between a circular domain (spot) and a
stripe in a square domain. This study can be performed analytically in
the limit of high degradation (furthermore, our study is exact in
$d=1$). The case of a finite degradation rate will be treated
numerically in another paper.

In previous papers \cite{gen,crrs,bose}, we have found a number of
analogies between the chemotactic problem and other systems of
physical interest (self-gravitating systems, 2D vortices,
Bose-Einstein condensation, Burgers equation). As mentioned
previously, the Keller-Segel model is isomorphic to the
Smoluchowski-Poisson system describing self-gravitating Brownian
particles \cite{crs}. In this analogy, the concentration of the
chemical produced by the bacteria plays a role similar to the
gravitational potential in astrophysical systems (they are both
solution of a Poisson equation) so that a number of analogies between
biology and gravity can be developped \cite{crrs}.  In addition, the
collapse of bacterial populations for $M>M_{c}$ or the collapse of
self-gravitating Brownian particles for $T<T_{c}$ is by many respects
similar to the Bose-Einstein condensation in phase space
\cite{bose}. Finally, there exists some analogies between the
chemotactic aggregation of bacteria and the formation of large-scale
vortices in 2D turbulence \cite{gen,crrs}. In that case, the
concentration of bacteria plays the role of the vorticity and the
concentration of the chemical produced by the bacteria plays the role
of the streamfunction. In two-dimensional hydrodynamics, the vorticity
field which is solution of the 2D Euler equation can achieve a
statistical equilibrium state (on the coarse-grained scale) as a
result of turbulent mixing (violent relaxation) \cite{houches}. In the
two-levels approximation, the equilibrium vorticity profile is
given by a Fermi-Dirac-like distribution
\cite{lb,miller,rs,csr}. Interestingly, this is similar to the steady state
of the regularized Keller-Segel model introduced in this
paper. Furthermore, in the quasi-geostrophic (Q.G.)  approximation
relevant to geophysical flows \cite{pedlosky}, the finite value of the Rossby
deformation radius introduces a shielding of the interaction between
vortices which is formally similar to the degradation of the chemical
in the chemotactic problem. In particular, the degradation rate $k$
plays the same role as the inverse of the deformation radius
$R^{-1}$. In the context of jovian vortices, Sommeria {\it et al.} 
\cite{nore} and Bouchet \& Sommeria \cite{bs} have considered the limit of a
small deformation radius $R\rightarrow 0$ to account for the annular
jet structure of Jupiter's great red spot. As we shall see, this is
similar to considering a limit of high degradation in the chemotactic
problem. Therefore, many interesting results can be obtained by
developing the analogies between these different topics.

The paper is organized as follows. In Sec. \ref{sec_reg}, we introduce
a regularized Keller-Segel model of chemotactic aggregation. We first
provide a phenomenological derivation of this model followed by a more
kinetic approach. In Sec. \ref{sec_dwall}, we study the equilibrium
states of this model in a limit of high degradation. For $T<T_{c}$, we
show that the solutions are formed by two phases in contact separated
by an interface (the stability of the uniform phase is considered in
Appendix \ref{sec_stab}). In Secs. \ref{sec_sts}-\ref{sec_wp}, we
develop a ``domain wall'' theory to study the properties of the
interface and determine its main characteristics (profile, width,
surface tension,...). Asymptotic behaviors of these expressions are
obtained for $T\rightarrow T_{c}$ and $T\rightarrow 0$ in
Secs. \ref{sec_tc} and \ref{sec_tz}. Analytical approximations of the
jet profile are given in Secs. \ref{sec_appsup} and
\ref{sec_appinf} in the form of self-similar solutions. Other
approximations are given in Sec. \ref{sec_ma} using match
asymptotics. In Sec. \ref{sec_curvature}, we show that the curvature
radius is constant so that, in two dimensions, the interface is either
a line (stripe) or a circle (spot). These results can be obtained
equivalently by minimizing the free energy functional associated with
the regularized Keller-Segel model (see Sec. \ref{sec_fe}). In
Sec. \ref{sec_bif}, we determine the phase diagram of the system and
the range of control parameters $(B,T)$ where the equilibrium state is
a stripe or a spot (the parameter $B$ is related to the total mass of
the configuration). Finally, in Sec. \ref{sec_analogy}, we develop the
close analogy between our biological system and the jet structure of
Jupiter's great red spot and other jovian vortices.

\section{The regularized Keller-Segel model}
\label{sec_reg}

\subsection{The dynamical equations}
\label{sec_dyn}

The general Keller-Segel model \cite{ks} describing the chemotaxis of bacterial populations consists in two coupled differential equations 
\begin{eqnarray}
\label{dyn1}
\frac{\partial\rho}{\partial t}=\nabla\cdot \left (D_{2}\nabla\rho)-\nabla\cdot (D_{1}\nabla c\right ),
\end{eqnarray}
\begin{eqnarray}
\label{dyn2}
\epsilon\frac{\partial c}{\partial t}=-k(c)c+f(c)\rho+D_{c}\Delta c, 
\end{eqnarray}
that govern the evolution of the density of bacteria $\rho({\bf r},t)$
and the evolution of the secreted chemical $c({\bf r},t)$. The
bacteria diffuse with a diffusion coefficient $D_{2}$ and they also
move in a direction of a positive gradient of the chemical
(chemotactic drift). The coefficient $D_{1}$ is a measure of the
strength of the influence of the chemical gradient on the flow of
bacteria. On the other hand, the chemical is produced by the bacteria
with a rate $f(c)$ and is degraded with a rate $k(c)$. It also
diffuses with a diffusion coefficient $D_{c}$. In the general
Keller-Segel model, $D_{1}=D_{1}(\rho,c)$ and $D_{2}=D_{2}(\rho,c)$
can both depend on the concentration of the bacteria and of the
chemical. This takes into account microscopic constraints, like
close-packing effects, that can hinder the movement of bacteria.

A very much studied version of the Keller-Segel model is provided by
the system of equations
\begin{eqnarray}
\label{dyn3}
\frac{\partial\rho}{\partial t}=\nabla\cdot \left (D\nabla\rho-\chi\rho\nabla c\right ),
\end{eqnarray}
\begin{eqnarray}
\label{dyn4}
\epsilon\frac{\partial c}{\partial t}=D'\Delta c+a\rho-bc,
\end{eqnarray}
where the parameters are positive constants.  Equation (\ref{dyn3}) can be
viewed as a mean-field Fokker-Planck equation associated with a Langevin
dynamics of the form
\begin{eqnarray}
\label{dyn5}
\frac{d{\bf r}}{dt}=\chi\nabla c+\sqrt{2D}{\bf R}(t),
\end{eqnarray}
where ${\bf R}(t)$ is a white noise and $\chi$ plays the role of a mobility. The stationary solution of Eq. (\ref{dyn3}) is given by
\begin{eqnarray}
\label{dyn6}
\rho=Ae^{\frac{\chi}{D}c}.
\end{eqnarray} 
This is similar to the Boltzmann distribution for a system in a
potential $-c$. This suggests to introducing an effective temperature
through the relation $T_{eff}=D/\chi$ which is similar to the Einstein
relation. For $\epsilon=0$, the system (\ref{dyn3})-(\ref{dyn4})
monotonically decreases ($\dot F\le 0$) the Lyapunov functional
\begin{eqnarray}
\label{dyn7}
F=-\frac{1}{2}\int \rho c \, d{\bf r}+\frac{D}{\chi}\int \rho\ln\rho \, d{\bf r},
\end{eqnarray} 
which is similar to a free energy $F=E-T_{eff}S$ where
$E=-\frac{1}{2}\int \rho c d{\bf r}$ is the energy of interaction and
$S=-\int \rho\ln\rho d{\bf r}$ is the Boltzmann entropic functional
\footnote{Note that these analogies with thermodynamics take even more
sense if we remark that the Keller-Segel model is isomorphic to the
Smoluchowski-Poisson system for self-gravitating Brownian particles
\cite{crrs}.}. For $\epsilon\neq 0$, the Lyapunov functional is
\begin{eqnarray}
\label{dyn8}
F=\frac{1}{2a}\int \left\lbrack D'(\nabla c)^{2}+b c^{2}\right \rbrack \, d{\bf r}-\int \rho c \, d{\bf r}+\frac{D}{\chi}\int \rho\ln\rho \, d{\bf r}.\nonumber\\
\end{eqnarray}

We shall consider here a more general situation where the mobility and
the diffusion coefficient in the Keller-Segel model can depend on the
density of bacteria. In that case, Eq. (\ref{dyn3}) is replaced by
\begin{eqnarray}
\label{dyn9}
\frac{\partial\rho}{\partial t}=\nabla\cdot \left \lbrack \nabla(D(\rho)\rho)-\chi(\rho)\rho\nabla c\right \rbrack.
\end{eqnarray} 
This can be viewed as a nonlinear mean-field Fokker-Planck equation
\cite{gen}. It is associated with a Langevin equation of the form
\begin{eqnarray}
\label{dyn10}
\frac{d{\bf r}}{dt}=\chi(\rho)\nabla c+\sqrt{2D(\rho)}{\bf R}(t).
\end{eqnarray}
We define the functions $h$ and $g$ by
\begin{eqnarray}
\label{dyn11}
Dh(\rho)=\frac{d}{d\rho}(\rho D(\rho)),
\end{eqnarray} 
\begin{eqnarray}
\label{dyn12}
\chi g(\rho)=\rho\chi(\rho),
\end{eqnarray} 
where $D$ and $\chi$ are positive coefficients. With these notations,
Eq. (\ref{dyn9}) can be rewritten
\begin{eqnarray}
\label{dyn13}
\frac{\partial\rho}{\partial t}=\nabla\cdot \left \lbrack Dh(\rho)\nabla\rho-\chi g(\rho)\nabla c\right \rbrack.
\end{eqnarray}
Setting $\beta=1/T_{eff}=\chi/D$, we obtain
\begin{eqnarray}
\label{dyn14}
\frac{\partial\rho}{\partial t}=\nabla\cdot \left \lbrack D\left (h(\rho)\nabla\rho-\beta  g(\rho)\nabla c\right )\right\rbrack.
\end{eqnarray}
This type of nonlinear mean-field Fokker-Planck equations has been
discussed in \cite{gen}. They are associated with generalized
entropic functionals of the form 
\begin{eqnarray}
\label{dyn14a}
S=-\int C(\rho)\, d{\bf r},
\end{eqnarray}
where $C(\rho)$ is a convex function defined by
\begin{eqnarray}
\label{dyn14b}
C''(\rho)=\frac{h(\rho)}{g(\rho)}.
\end{eqnarray}

The Keller-Segel model (\ref{dyn3})-(\ref{dyn4}) is known to exhibit
blow-up solutions when the chemotactic attraction prevails over
diffusion \cite{horstmann,banach}. This reproduces the chemotactic
aggregation of bacterial populations. In theory, the density can take
arbitrarily large values and ultimately forms a Dirac peak. In
reality, this singular evolution is unphysical as we expect finite
size effects and close-packing effects to become important when the
system aggregates and becomes dense enough. We shall regularize the
problem by introducing a sort of filling factor in the drift-diffusion
equation (\ref{dyn14}). Thus, we take $h(\rho)=1$ and
$g(\rho)=\rho(1-\rho/\sigma_{0})$ so that Eq. (\ref{dyn3}) is replaced by 
\begin{eqnarray}
\label{dyn15}
\frac{\partial\rho}{\partial t}=\nabla\cdot \left \lbrack D\left (\nabla\rho-\beta\rho(1-\rho/\sigma_{0})\nabla c\right )\right \rbrack.
\end{eqnarray}
With this modification, the mobility is reduced when the density
becomes high enough (i.e. when $\rho$ approaches the value
$\sigma_{0}$) and this prevents singularities to form. Indeed, it can
be shown that the density remains always bounded: $\rho({\bf r},t)\le
\sigma_{0}$. This bound is similar to the Pauli exclusion principle in 
quantum mechanics, but it occurs here in physical space. The
regularized drift-diffusion equation (\ref{dyn15}) was introduced
phenomenologically in \cite{gen,crrs} to avoid infinite values of the
density.

In the following, we shall take $\epsilon=0$ for simplicity (although
many results which are valid at equilibrium are independent on this
assumption). This is valid in a limit of high diffusivity of the
chemical \cite{jager}.  If we introduce the notations $k^{2}=b/D'$ and
$\lambda=a/D'$, we obtain the regularized Keller-Segel model
\begin{eqnarray}
\label{dyn16}
\frac{\partial\rho}{\partial t}=\nabla\cdot \left \lbrack D\left (\nabla\rho-\beta \rho(1-\rho/\sigma_{0})\nabla c\right )\right \rbrack,
\end{eqnarray}
\begin{eqnarray}
\label{dyn17}
\Delta c-k^{2}c=-\lambda\rho.
\end{eqnarray}
The stationary solution of Eq. (\ref{dyn16}) is given by
\begin{eqnarray}
\label{dyn18}
\rho=\frac{\sigma_{0}}{1+e^{-\beta c+\alpha}},
\end{eqnarray}
which is similar to the Fermi-Dirac distribution in physical
space. From this expression, we clearly have $\rho({\bf r})\le
\sigma_{0}$. Furthermore, the Lyapunov functional can be written 
in the form of a
free energy $F=E-T_{eff}S$ where
\begin{eqnarray}
\label{dyn19}
E=-\frac{1}{2}\int \rho c \, d{\bf r}=-\frac{1}{2\lambda}\int \left\lbrack (\nabla c)^{2}+k^{2}c^{2}\right\rbrack \, d{\bf r},
\end{eqnarray}
is the energy of interaction and 
\begin{equation}
\label{dyn20}
S=-\sigma_{0}\int \left\lbrace \frac{\rho}{\sigma_{0}}\ln\frac{\rho}{\sigma_{0}}+\left (1-\frac{\rho}{\sigma_{0}}\right )\ln\left (1-\frac{\rho}{\sigma_{0}}\right )\right\rbrace d{\bf r},  
\end{equation}
is the Fermi-Dirac entropic functional in physical space. The
distribution (\ref{dyn18}) extremizes the free energy at fixed
mass. Indeed, writing the first order variations in the form $\delta
F+\alpha T_{eff}\delta M=0$ where $\alpha$ is a Lagrange multiplier,
we recover Eq. (\ref{dyn18}). Furthermore, it can be shown that a
stationary solution of Eqs. (\ref{dyn16})-(\ref{dyn17}) is linearly
stable if, and only if, it is a {\it minimum} of $F$ at fixed mass \cite{gen}.

\subsection{Phenomenological derivation of the model}
\label{sec_phen}

In this section, we develop a connection between the chemotactic
problem and thermodynamics. To that purpose, we introduce the entropic
functional (\ref{dyn20}) from a combinatorial analysis which respects
an exclusion principle in physical space. Then, we obtain the
dynamical equation (\ref{dyn16}) from arguments similar to the linear
thermodynamics of Onsager.

We divide the domain into a very large number of microcells with size
$h$. We assume that the size $h$ is of the order of the size of a
particle so that a microcell is occupied either by $0$ or $1$
particle. This is how the exclusion principle is introduced in the
problem. We shall now group these microcells into macrocells each of
which contains many microcells but remains nevertheless small compared
to the spatial extension of the whole system. We call $\nu$ the number
of microcells in a macrocell. Consider the configuration $\lbrace n_i
\rbrace$ where there are $n_1$ particles in the $1^{\rm st}$
macrocell, $n_2$ in the $2^{\rm nd}$ macrocell etc..., each occupying
one of the $\nu$ microcells with no cohabitation. The number of ways
of assigning a microcell to the first element of a macrocell is $\nu$,
to the second $\nu -1$ etc. Assuming that the particles are
indistinguishable, the number of ways of assigning microcells to all
$n_i$ particles in a macrocell is thus
\begin{equation}
{1\over n_i!}{\times} {\nu!\over (\nu-n_i)!}. \label{phen1}
\end{equation}
To obtain the number of microstates corresponding to the macrostate
$\lbrace n_i \rbrace$ defined by the number of particles $n_i$ in each
macrocell (irrespective of their precise position in the cell), we
need to take the product of terms such as (\ref{phen1}) over all
macrocells. Thus, the number of microstates corresponding to the
macrostate $\lbrace n_i \rbrace$, which is proportional to the {\it a
priori} probability of the state $\lbrace n_i \rbrace$, is
\begin{equation}
W(\lbrace n_i \rbrace)=\prod_i {\nu!\over n_i!(\nu-n_i)!}.
\label{phen2}
\end{equation}
This is the Fermi-Dirac statistics which is applied here in physical
space. As is customary, we define the entropy of the state $\lbrace
n_i \rbrace$ by
\begin{equation}
S(\lbrace n_i \rbrace)=\ln W(\lbrace n_i \rbrace). \label{phen3}
\end{equation}
It is convenient here to return to a representation in terms of
the density in the
$i$-th macrocell
\begin{equation}
\rho_i=\rho({\bf r}_i)={n_i \ m\over \nu h^d}={n_i\sigma_0\over \nu}, \label{phen4}
\end{equation}
where we have defined $\sigma_0=m/h^d$, which represents the
maximum value of $\rho$ due to the exclusion constraint. Now,
using the Stirling formula, we have
\begin{eqnarray}
\ln W(\lbrace n_i \rbrace)\simeq \sum_i \nu
(\ln\nu-1)-\nu\biggl\lbrace {\rho_i\over \sigma_0}\biggl\lbrack
\ln\biggl ({\nu \rho_i\over \sigma_0}\biggr )-1\biggr\rbrack \nonumber\\
 +\biggl
(1-{\rho_i\over \sigma_0}\biggr )\biggl\lbrack\ln\biggl\lbrace
\nu\biggl (1-{\rho_i\over \sigma_0}\biggr
)\biggr\rbrace-1\biggr\rbrack\biggr\rbrace.\qquad  \label{phen5}
\end{eqnarray}
Passing to the continuum limit $\nu\rightarrow 0$, we obtain the
expression (\ref{dyn19}) of the Fermi-Dirac entropy in physical
space. In the dilute limit $\rho\ll \sigma_{0}$, it reduces to the
Boltzmann entropy (\ref{dyn7}).

The entropy is the correct thermodynamical potential for an isolated
system for which the energy is conserved (microcanonical
ensemble). This is not the case for our system which is
dissipative. The proper description is the canonical ensemble and the
correct thermodynamical potential is the free energy $F=E-T_{eff}S$
constructed with the Fermi-Dirac entropy (\ref{dyn20}) and the energy
(\ref{dyn19}). The equilibrium state in the canonical ensemble is
obtained by minimizing the free energy at fixed mass. Writing the
first variations as $\delta F-\lambda\delta M=0$, we obtain
\begin{eqnarray}
\label{phen6}
\frac{\delta F}{\delta\rho}-\lambda=0,
\end{eqnarray}
which leads to the Fermi-Dirac distribution (\ref{dyn18}) with
$\alpha=-\lambda\beta$. Now that the proper thermodynamical potential
has been derived by a combinatorial analysis, we can introduce
phenomenologically a dynamical model by writing the evolution of the
density as a continuity equation $\partial_{t}\rho=\nabla \cdot {\bf
J}$ where the current is the gradient of the functional
derivative of the free energy, i.e.
\begin{eqnarray}
\label{phen7}
\frac{\partial\rho}{\partial t}=\nabla\cdot \left (\mu\nabla\frac{\delta F}{\delta\rho}\right ).
\end{eqnarray}
This formulation ensures that the free energy decreases monotonically provided that $\mu\ge 0$. Indeed,
\begin{eqnarray}
\label{phen8}
\dot F=\int \frac{\delta F}{\delta\rho}\frac{\partial\rho}{\partial t}\, d{\bf r}=\int \frac{\delta F}{\delta\rho} \nabla\cdot {\bf J}\, d{\bf r}\nonumber\\
=-\int {\bf J}\cdot \nabla \frac{\delta F}{\delta\rho}\, d{\bf r}=-\int \mu\left (\nabla \frac{\delta F}{\delta\rho}\right )^{2}\, d{\bf r}\le 0.
\end{eqnarray}
Furthermore, a steady state satisfies $\dot F=0$, i.e $\nabla (\delta F/\delta\rho)=0$ leading to Eq. (\ref{phen6}).  Now, using Eqs. (\ref{dyn19}) and
(\ref{dyn20}), we have
\begin{eqnarray}
\label{phen9}
\nabla \frac{\delta F}{\delta\rho}=-\nabla c+\frac{D}{\chi}\frac{\nabla\rho}{\rho (1-\rho/\sigma_{0})}.
\end{eqnarray}
To avoid the singularity when $\rho=0$ or $\rho=\sigma_{0}$, we require that $\mu$ is proportional to $\rho (1-\rho/\sigma_{0})$. Writing Eq. (\ref{phen7}) in the form
\begin{eqnarray}
\label{phen10}
\frac{\partial\rho}{\partial t}=\nabla\cdot \left \lbrack \chi\rho (1-\rho/\sigma_{0})\nabla\frac{\delta F}{\delta\rho}\right \rbrack,
\end{eqnarray}
and using Eq. (\ref{phen9}), we obtain Eq. (\ref{dyn15}). This
approach to construct relaxation equations is equivalent to
Onsager's linear thermodynamics. Indeed, noting that the potential
\begin{eqnarray}
\label{phen11}
\lambda({\bf r},t)\equiv \frac{\delta F}{\delta\rho}=- c+T_{eff}\ln\left (\frac{\rho/\sigma_{0}}{1-\rho/\sigma_{0}}\right ),
\end{eqnarray}
is uniform at equilibrium according to Eq. (\ref{dyn18}) or (\ref{phen6}), the linear thermodynamics of Onsager suggests to writing the  current as
\begin{eqnarray}
\label{phen12}
{\bf J}=\mu\nabla\lambda({\bf r},t),
\end{eqnarray}
which is equivalent to Eq. (\ref{phen7}). The same results can be
obtained by a variational formulation which is related to the Maximum
Entropy Production Principle (MEPP) \cite{gen}. The rate of
dissipation of free energy is given by
\begin{eqnarray}
\label{phen13}
\dot F=\int \frac{\delta F}{\delta\rho}\frac{\partial\rho}{\partial t}\, d{\bf r}=\int \frac{\delta F}{\delta\rho} \nabla\cdot {\bf J}\, d{\bf r}
=-\int {\bf J}\cdot \nabla \frac{\delta F}{\delta\rho}\, d{\bf r}.\nonumber\\
\end{eqnarray}
We shall determine the optimal current ${\bf J}_{*}$ which maximizes
the rate of dissipation of free energy $\dot F$ under the constraint
$J^{2}\le C({\bf r},t)$ putting a (physical) bound on $|{\bf J}|$. The
corresponding variational problem can be written
\begin{eqnarray}
\label{phen14}
\delta \dot F+\delta\int \frac{{\bf J}^{2}}{2\mu}\, d{\bf r}=0,
\end{eqnarray}
where $\mu$ is a Lagrange multiplier. Performing the variations on ${\bf J}$, we obtain ${\bf
J}_{*}=\mu\nabla (\delta F/\delta\rho)$ which returns
Eq. (\ref{phen7}).

\subsection{Kinetic derivation of the model}
\label{sec_kin}

As discussed previously, Eq. (\ref{dyn14}) can be viewed as a
nonlinear Fokker-Planck equation where the diffusion coefficient and
the mobility explicitly depend on the local concentration of
particles. Such generalized Fokker-Planck equations can be derived
from a kinetic theory, starting from the master equation, and assuming
that the probabilities of transition explicitly depend on the
occupation number (concentration) of the initial and arrival
states. Below, we briefly summarize and adapt to the present situation
the approach carried out by Kaniadakis \cite{kaniadakis} in a more
general context.

We introduce a stochastic dynamics by defining the probability of transition of a bacteria from position ${\bf r}$ to position ${\bf r}'$. Following Kaniadakis \cite{kaniadakis}, we assume the following form  
\begin{eqnarray}
\label{kin1}
\pi({\bf r}\rightarrow {\bf r}')=w({\bf r},{\bf r}-{\bf r}')a\lbrack\rho({\bf r},t)\rbrack b\lbrack\rho({\bf r}',t)\rbrack.
\end{eqnarray}
Usual stochastic processes correspond to $a(\rho)=\rho$ and
$b(\rho)=1$: the probability of transition is proportional to the
density of the initial state and independent on the density of the
final state.  They lead to the Fokker-Planck equation (\ref{dyn3}) as will be
shown below. Here, we assume a more general dependence on the
occupancy in the initial and arrival states. This can account for
microscopic constraints like close-packing effects that can inhibitate
the transition. Quite generally, the evolution of the density
satisfies the master equation
\begin{eqnarray}
\label{kin2}
\frac{\partial\rho}{\partial t}=\int \left\lbrack \pi({\bf r}'\rightarrow {\bf r})-\pi({\bf r}\rightarrow {\bf r}')\right\rbrack d{\bf r}'.
\end{eqnarray}
Assuming that the evolution is sufficiently slow, and local, such that the dynamics only permits values of ${\bf r}'$ close to ${\bf r}$, one can develop the term in brackets in Eq. (\ref{kin2}) in powers of  ${\bf r}-{\bf r}'$. Proceeding along the lines of \cite{kaniadakis}, we obtain a Fokker-Planck-like equation 
\begin{equation}
\label{kin3}
\frac{\partial\rho}{\partial t}=\frac{\partial}{\partial x_{i}}\left\lbrack\left (\zeta_{i}+\frac{\partial\zeta_{ij}}{\partial x_{j}}\right )\gamma(\rho)+\gamma(\rho)\frac{\partial\ln \kappa(\rho)}{\partial\rho}\zeta_{ij}\frac{\partial\rho}{\partial x_{j}}\right\rbrack,
\end{equation}
with
\begin{equation}
\label{kin4}
\gamma(\rho)=a(\rho)b(\rho),\qquad \kappa(\rho)=\frac{a(\rho)}{b(\rho)},
\end{equation}
and
\begin{equation}
\label{kin5}
\zeta_{i}({\bf r})=-\int y_{i}w({\bf r},{\bf y})d{\bf y},
\end{equation}
\begin{equation}
\label{kin6}
\zeta_{ij}({\bf r})=\frac{1}{2}\int y_{i}y_{j}w({\bf r},{\bf y})d{\bf y}.
\end{equation}
The moments $\zeta_{i}$ and $\zeta_{ij}$ are fixed by the Langevin equations (\ref{dyn5}). Assuming isotropy
\begin{equation}
\label{kin7}
\zeta_{i}=J_{i}, \qquad \zeta_{ij}=D\delta_{ij},
\end{equation}
the kinetic equation becomes
\begin{equation}
\label{kin8}
\frac{\partial\rho}{\partial t}=\nabla\cdot \left\lbrack ({\bf J}+\nabla D)\gamma(\rho)+\gamma(\rho)\frac{\partial\ln \kappa(\rho)}{\partial\rho}D\nabla \rho\right\rbrack.
\end{equation}
Now, according to the Langevin equations (\ref{dyn5}), $D$ is independent on ${\bf r}$ and ${\bf J}=-\chi\nabla c$. Thus, we get
\begin{equation}
\label{kin9}
\frac{\partial\rho}{\partial t}=\nabla\cdot \left\lbrack D\gamma(\rho)\frac{\partial\ln \kappa(\rho)}{\partial\rho}\nabla\rho-\chi\gamma(\rho)\nabla c \right\rbrack.
\end{equation}
If we define
\begin{equation}
\label{kin10}
h(\rho)=\gamma(\rho)\frac{\partial\ln \kappa(\rho)}{\partial\rho}, \qquad g(\rho)=\gamma(\rho),
\end{equation}
the foregoing equation can be written
\begin{eqnarray}
\label{kin11}
\frac{\partial\rho}{\partial t}=\nabla\cdot \left \lbrack Dh(\rho)\nabla\rho-\chi g(\rho)\nabla c\right \rbrack,
\end{eqnarray}
and it coincides with the phenomenological equation (\ref{dyn13}). It
seems natural to assume that the transition probability is
proportional to the density of the initial state so that
$a(\rho)=\rho$. In that case, we obtain an equation of the form
\begin{equation}
\label{kin12}
\frac{\partial\rho}{\partial t}=\nabla\cdot \left ( D\left\lbrack b(\rho)-\rho b'(\rho)\right \rbrack \nabla\rho-\chi\rho b(\rho)\nabla c\right ). 
\end{equation}
Note that the coefficients of diffusion and mobility are not
independent since they are both expressed in terms of
$b(\rho)$. Choosing $b(\rho)=1$, i.e. a probability of transition which
does not depend on the population of the arrival state, leads to the
standard Fokker-Planck equation (\ref{dyn3}).  If, now, we assume that
the transition probability is blocked (inhibited) if the concentration
of the arrival state is equal to $\sigma_0$, then it seems natural to
take $b(\rho)=1-\rho/\sigma_{0}$. In that case, we obtain
\begin{equation}
\label{kin13}
\frac{\partial\rho}{\partial t}=\nabla\cdot \left ( D\nabla\rho-\chi\rho (1-\rho/\sigma_{0})\nabla c\right ),
\end{equation}
which coincides with the phenomenological equation (\ref{dyn15}).

We can consider a related kinetic model with similar thermodynamical
properties and the same equilibrium states. For this model, the dynamical
equation reads
\begin{equation}
\label{kin14}
\frac{\partial\rho}{\partial t}=\nabla\cdot \left \lbrack \chi\left (\frac{T_{eff}}{1-\rho/\sigma_{0}}\nabla\rho-\rho\nabla c\right )\right\rbrack.
\end{equation}
This can be put in the form of a generalized Smoluchowski equation \cite{gen}:
\begin{equation}
\label{kin15}
\frac{\partial\rho}{\partial t}=\nabla\cdot \left \lbrack \frac{1}{\xi}\left (\nabla p-\rho\nabla c\right )\right\rbrack,
\end{equation}
associated with a barotropic equation of state
$p(\rho)=-\sigma_{0}T_{eff}\ln(1-\rho/\sigma_{0})$, where $p$ is an
effective ``pressure''. For $\rho\ll\sigma_{0}$, we recover the
``isothermal'' equation of state $p=\rho T_{eff}$ leading to
the ordinary Keller-Segel model (\ref{dyn3}).

Equations (\ref{kin13}) and (\ref{kin14}) have very similar
properties and they can be viewed as natural extensions of the Keller-Segel model. In Eq. (\ref{kin13}) the regularization is put in the drift term
while in Eq. (\ref{kin14}) it is put in the diffusion term. These two
possibilities are considered in
\cite{gen}. Note finally that Eq. (\ref{kin14}) can be obtained from the master equation (\ref{kin2}) when the transition
probabilities are of the form (\ref{kin1}) with
$a(\rho)=\rho/\sqrt{1-\rho/\sigma_{0}}$ and
$b(\rho)=\sqrt{1-\rho/\sigma_{0}}$.

\section{Domain wall theory}
\label{sec_dwall}

\subsection{The stationary state}
\label{sec_sts}

The stationary solution of the regularized Keller-Segel model
(\ref{dyn16}) is the Fermi-Dirac-like distribution
\begin{eqnarray}
\label{sts1}
\rho=\frac{\sigma_{0}}{1+e^{-\beta c+\alpha}}.
\end{eqnarray}
This relates, at equilibrium, the bacterial density $\rho$ to the
concentration of the chemical $c$. The chemical is itself produced by
the bacteria according to Eq. (\ref{dyn17}). Thus, combining
Eqs. (\ref{sts1}) and (\ref{dyn17}), we obtain a differential
equation for the concentration $c$.  Using the identity
\begin{eqnarray}
\label{sts2}
\frac{1}{1+e^{x}}=\frac{1}{2}\left \lbrack 1-\tanh \left (\frac{x}{2}\right )\right \rbrack,
\end{eqnarray}
this  mean-field equation can be written
\begin{eqnarray}
\label{sts3}
\Delta c-k^{2}c=-\frac{\lambda \sigma_{0}}{2}\left \lbrack 1-\tanh \left (\frac{\alpha-\beta c}{2}\right )\right \rbrack.
\end{eqnarray}
Introducing the new variables
\begin{eqnarray}
\label{sts4}
\psi=\frac{c}{\left ({\lambda\sigma_{0}}/{2}\right )}, \quad \mu=\frac{\alpha}{2}, \quad C=\frac{\beta \lambda\sigma_{0}}{4k^{2}},
\end{eqnarray}
we get
\begin{eqnarray}
\label{sts5}
\Delta \psi-k^{2}\psi=-1+\tanh \left \lbrack C\left (\frac{\mu}{C}-k^{2}\psi\right )\right\rbrack.
\end{eqnarray}
We shall solve this equation perturbatively in a limit of high degradation so that $k\gg 1$ (note that the following treatment is exact in $1D$ without approximation). In that limit, we can neglect the gradient of concentration except in a thin layer of length $\sim k^{-1}$ (domain wall) where the concentration changes rapidly. Outside the wall, we obtain the algebraic equation
\begin{eqnarray}
\label{sts6}
-k^{2}\psi=-1+\tanh \left \lbrack C\left (\frac{\mu}{C}-k^{2}\psi\right )\right\rbrack.
\end{eqnarray}
In some range of parameters (see below), this equation determines two
solutions $\psi_{\pm}$ which correspond to two phases with uniform
concentration. These two phases are connected by a ``wall''. For $k\gg
1$, the interface is very thin so that we can neglect the
curvature of the wall in a first approximation.  The wall profile is
then determined by the one-dimensional differential equation
\begin{eqnarray}
\label{sts7}
\frac{d^{2}\psi}{d\xi^{2}}-k^{2}\psi=-1+\tanh \left \lbrack C\left (\frac{\mu}{C}-k^{2}\psi\right )\right\rbrack,
\end{eqnarray}
where $\xi$ is a coordinate normal to the interface.

\subsection{The wall equation}
\label{sec_we}

If we set
\begin{eqnarray}
\label{we1}
\phi=k^{2}\psi-\mu/C, \quad {\tau}=k{\xi}, \quad \chi=\frac{\mu}{C}-1,
\end{eqnarray}
we obtain the wall equation 
\begin{eqnarray}
\label{we2}
\frac{d^{2}\phi}{d\tau^{2}}=-\tanh (C\phi)+\phi+\chi. 
\end{eqnarray}
Eq. (\ref{we2}) is similar to the equation of motion for a particle in
a potential
\begin{eqnarray}
\label{we3}
U(\phi)=T\ln\left \lbrack \cosh (\phi/T)\right\rbrack-\frac{\phi^{2}}{2}+\chi\phi+U_{0}, 
\end{eqnarray}
where $\tau$ plays the role of time and $\phi$ plays the role of
position.  Indeed, it can be rewritten
\begin{eqnarray}
\label{we4}
\frac{d^{2}\phi}{d\tau^{2}}=-U'(\phi).
\end{eqnarray}
We have introduced the notation $C=1/T$ where, as we shall see, $T$
plays the role of a temperature (it is furthermore proportional to the
effective temperature $T_{eff}$). On the other hand, $U_{0}$ is a
constant of integration that will be specified later.

\begin{figure}
\centering
\includegraphics[width=8cm]{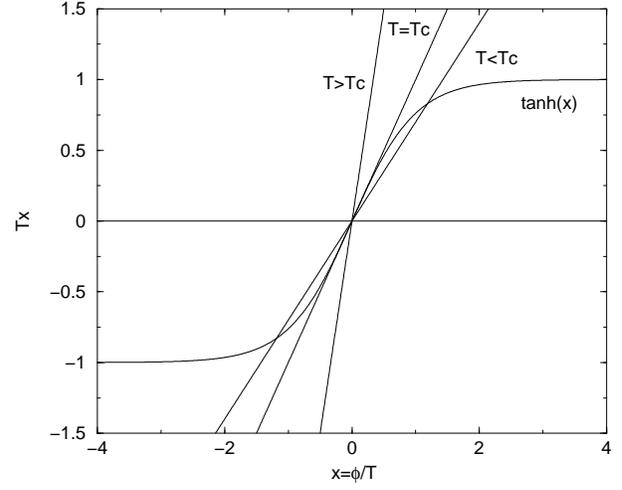}
\caption{Graphical construction determining the solutions $\pm u$ of the algebraic equation (\ref{we10})-b as a function of the temperature $T$. }
\label{tanh}
\end{figure}

\begin{figure}
\centering
\includegraphics[width=8cm]{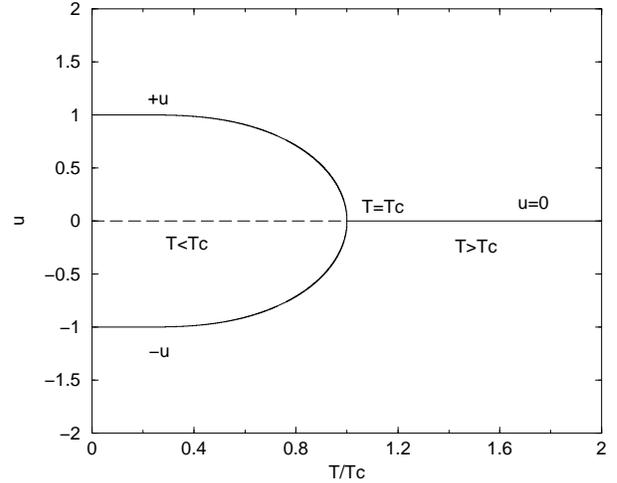}
\caption{Evolution of the order parameter $u$ as a function of the temperature $T$. For $T>T_{c}$ the system is homogeneous with $\phi=u=0$. For $T<T_{c}$, the uniform solution becomes unstable (see Appendix \ref{sec_stab}) and  two phases  $\phi_{\pm}=\pm u$ separated by a ``domain wall'' appear.   }
\label{utnew}
\end{figure}

Far from the wall, the density is uniform with values $\phi_{\pm}$ satisfying
\begin{eqnarray}
\label{we5}
U'(\phi_{\pm})=0.
\end{eqnarray} 
On the other hand, a first integral of Eq. (\ref{we4}) is
\begin{eqnarray}
\label{we6}
E=\frac{1}{2}\left (\frac{d\phi}{d\tau}\right )^{2}+U(\phi).
\end{eqnarray} 
The condition of solvability is therefore
\begin{eqnarray}
\label{we7}
U(\phi_{-})=U(\phi_{+}).
\end{eqnarray} 
The only possibility to satisfy the two conditions
(\ref{we5})-(\ref{we7}) simultaneously is that $\chi=0$, i.e
$\mu=\frac{1}{T}$ (in that case, $U(\phi)$ is symmetric and the above
conditions are satisfied trivially).  Then, the wall profile is
completely determined by the equations
\begin{eqnarray}
\label{we8}
\frac{d^{2}\phi}{d\tau^{2}}=-\tanh (\phi/T)+\phi,
\end{eqnarray}
\begin{eqnarray}
\label{we9}
U(\phi)=T\ln\left \lbrack \cosh (\phi/T)\right\rbrack
-\frac{\phi^{2}}{2}+U_{0}, 
\end{eqnarray}
\begin{eqnarray}
\label{we10}
\phi_{\pm}=\pm u, \qquad u=\tanh(u/T).
\end{eqnarray}
We note that the algebraic equation has solutions $u\neq 0$ only if (see Fig. \ref{tanh})
\begin{eqnarray}
\label{we11}
T<T_{c}=1.
\end{eqnarray}
This is similar to a second order phase transition (see
Fig. \ref{utnew}). In that case, the algebraic equation has three
solutions $0$ and $\pm u$ with $u\le 1$ (but the solution $\phi=u=0$
is unstable). The concentrations of bacteria and chemical in the
uniform domains are related to the order parameter $u$ by
\begin{eqnarray}
\label{we12}
c_{\pm}=\frac{\lambda \sigma_{0}}{2k^{2}}(1\pm u),\quad \rho_{\pm}=\frac{\sigma_{0}}{2}(1\pm u).
\end{eqnarray}
More generally, using Eqs. (\ref{sts1})-(\ref{sts2}) the concentration
profiles in the whole space can be expressed in terms of the field
$\phi$ by
\begin{eqnarray}
\label{we13}
c=\frac{\lambda \sigma_{0}}{2k^{2}}(1+\phi),\quad \rho=\frac{\sigma_{0}}{2}\left\lbrack 1-\tanh(\phi/T)\right \rbrack.
\end{eqnarray}

\subsection{The wall profile}
\label{sec_wp}

\begin{figure}
\centering
\includegraphics[width=8cm]{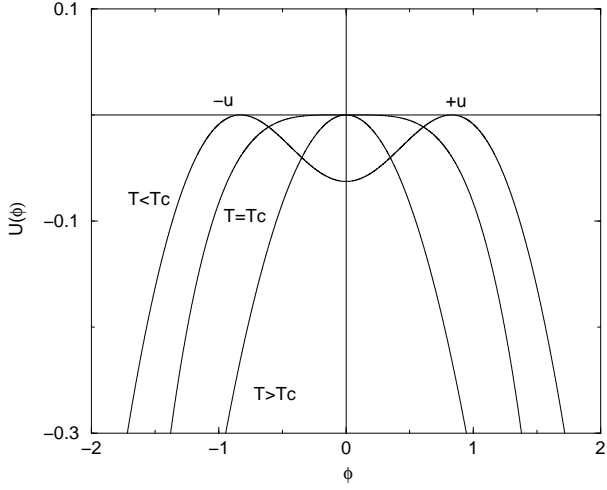}
\caption{The potential $U(\phi)$ of the equivalent mechanical problem for different values of the temperature.}
\label{Uphi}
\end{figure}

\begin{figure}
\centering
\includegraphics[width=8cm]{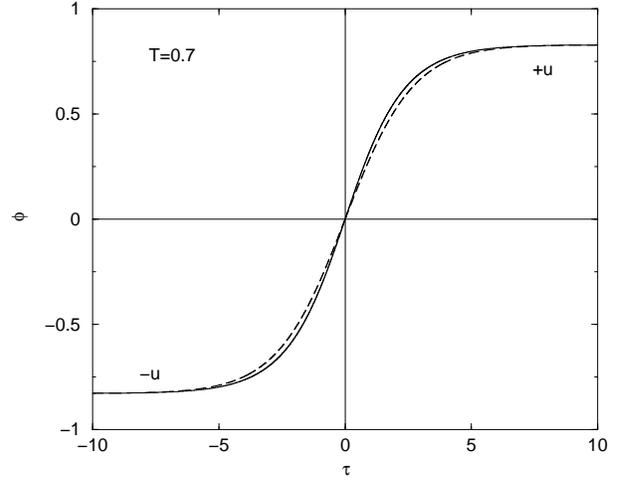}
\caption{Plot of the field $\phi$ accross the wall for $T=0.7$. The solid line corresponds to the exact solution of Eq. (\ref{wp2}) obtained numerically and the dashed line corresponds to the approximate expression (\ref{appsup1}).   }
\label{phitau7}
\end{figure}

\begin{figure}
\centering
\includegraphics[width=8cm]{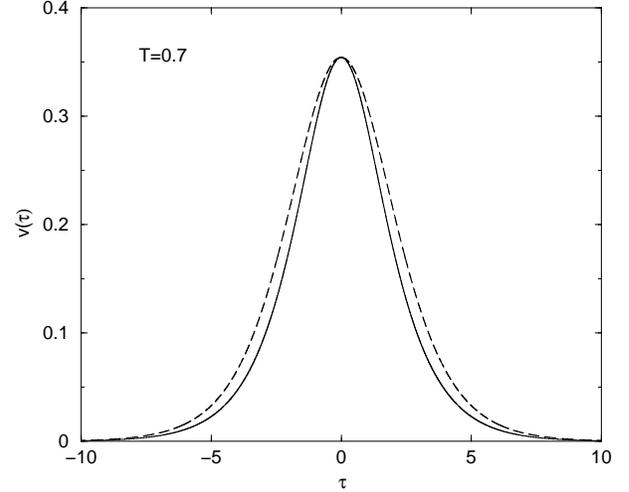}
\caption{Plot of the concentration gradient  $v(\tau)$ accross the wall for $T=0.7$. The solid line corresponds to the exact solution of Eq. (\ref{wp2}) obtained numerically and the dashed line corresponds to the approximate expression (\ref{appsup4})-(\ref{appsup6}). }
\label{tauV7}
\end{figure}

\begin{figure}
\centering
\includegraphics[width=8cm]{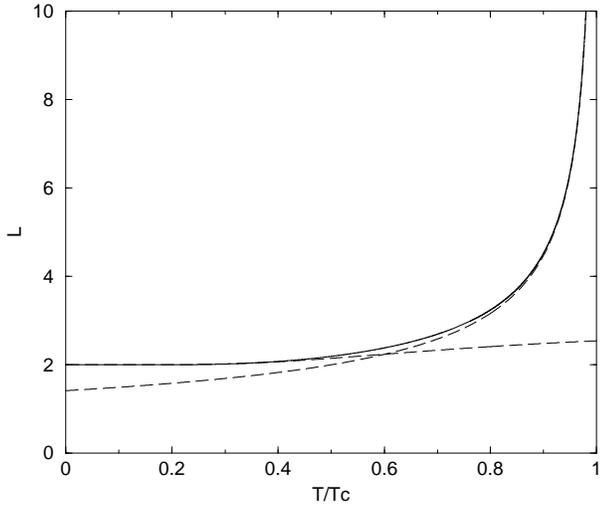}
\caption{Evolution of the typical width of the wall as a function of the temperature. The dashed lines correspond to the asymptotic expressions for $T\rightarrow 0$ and $T\rightarrow T_{c}$.  }
\label{LT}
\end{figure}

\begin{figure}
\centering
\includegraphics[width=8cm]{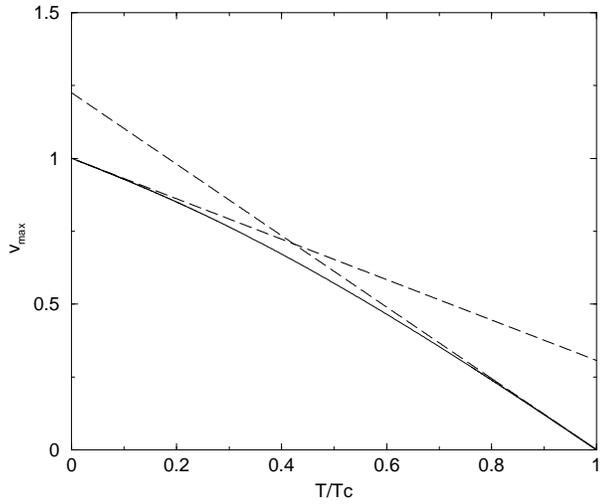}
\caption{Evolution of the maximum concentration gradient as a function of the temperature. The dashed lines correspond to the asymptotic expressions for $T\rightarrow 0$ and $T\rightarrow T_{c}$. }
\label{TV}
\end{figure}

\begin{figure}
\centering
\includegraphics[width=8cm]{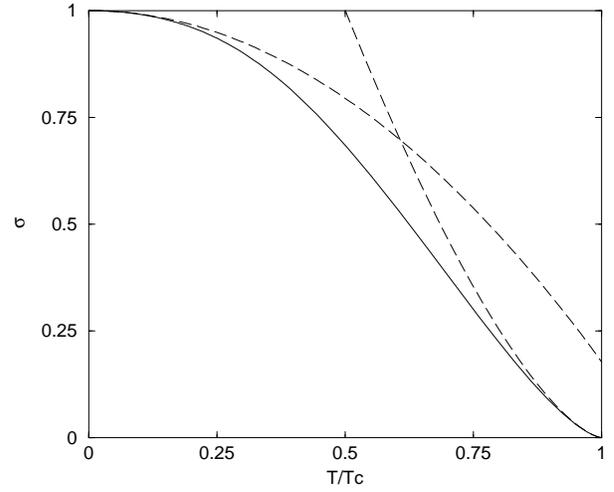}
\caption{Evolution of the energy of the wall (surface tension) as a function of the temperature. The dashed lines correspond to the asymptotic expressions for $T\rightarrow 0$ and $T\rightarrow T_{c}$. }
\label{tensionNEW}
\end{figure}

\begin{figure}
\centering
\includegraphics[width=8cm]{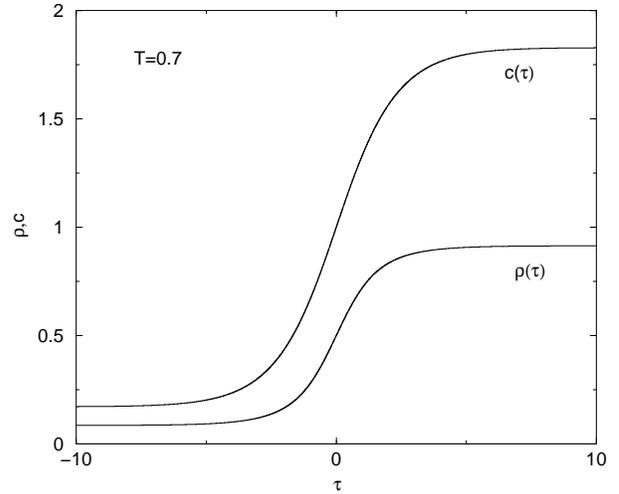}
\caption{Concentration profiles of bacteria $\rho$ (in units of $\sigma_{0}$) and of the secreted chemical $c$ (in units of $\lambda\sigma_{0}/2k^{2}$) for $T=0.7$.   }
\label{crho7}
\end{figure}

We determine $U_{0}$ such that $U(u)=0$. The potential is then
explicitly given by (see Fig. \ref{Uphi}):
\begin{equation}
\label{wp1}
U(\phi)=T\ln\left \lbrack \cosh \left (\frac{\phi}{T}\right )\right\rbrack-\frac{\phi^{2}}{2}-T\ln\left \lbrack \cosh \left (\frac{u}{T}\right )\right\rbrack+\frac{u^{2}}{2}. 
\end{equation} 
With this convention, the constant appearing in Eq. (\ref{we6}) is
$E=0$. Then, we obtain the equation
\begin{eqnarray}
\label{wp2}
\frac{1}{2}\left (\frac{d\phi}{d\tau}\right )^{2}=-U(\phi),
\end{eqnarray} 
which determines the wall profile by a simple integration (see Fig. \ref{phitau7})
\begin{eqnarray}
\label{wp3}
\int_{0}^{\phi}{dx\over \sqrt{-2U(x)}}=\tau.
\end{eqnarray} 
For $\tau\rightarrow +\infty$, $\phi\rightarrow u$. To get the asymptotic behaviour, we set $\phi=u-\theta$ with $\theta\ll 1$ and we linearize the wall equation (\ref{wp2}). This yields
\begin{eqnarray}
\label{wp4}
\frac{d\theta}{d\tau}=-\sqrt{-U''(u)}\theta,
\end{eqnarray} 
with $U''(u)={1\over T}(1-u^{2})-1<0$ for $T<T_{c}$. The wall connects the uniform phase exponentially rapidly. Thus, we can write
\begin{eqnarray}
\label{wp5}
\phi=u-A(u)e^{-2\tau/L(u)}, \qquad (\tau\rightarrow +\infty)
\end{eqnarray}
where the typical width of the wall (expressed in units of $k^{-1}$) is
\begin{eqnarray}
\label{wp6}
L(u)=\frac{2}{\sqrt{-U''(u)}}=\frac{2}{\sqrt{1-\frac{1}{T}(1-u^{2})}}.
\end{eqnarray}
We introduce the concentration gradient (see Fig. \ref{tauV7})
\begin{eqnarray}
\label{wp7}
v(\tau)=\frac{d\phi}{d\tau}=\sqrt{-2U(\phi)}.
\end{eqnarray} 
Using Eq. (\ref{wp1}), we find that the maximum value of the concentration gradient, corresponding to $\phi=0$, is given by
\begin{equation}
\label{wp8}
v_{max}(u)=\sqrt{2T\ln\left \lbrack \cosh \left (\frac{u}{T}\right )\right\rbrack-u^{2}}.
\end{equation} 
Finally, the energy of the wall, or surface tension, is
\begin{eqnarray}
\label{wp9}
\sigma(u)=\int_{-\infty}^{+\infty}\left (\frac{d\phi}{d\tau}\right )^{2}d\tau=\int_{-u}^{+u}\sqrt{-2U(\phi)}d\phi.
\end{eqnarray} 
The functions $L$, $\sigma$ and $v_{max}$ are plotted in
Figs. \ref{LT}, \ref{TV} and \ref{tensionNEW} as a function of the
temperature $T$, together with their asymptotic expressions computed in
the following sections. The concentration profiles of bacteria and of the secreted chemical, given by Eq. (\ref{we13}), are plotted in Fig. \ref{crho7}.

\subsection{The limit $T\rightarrow T_{c}$}
\label{sec_tc}

For $T\rightarrow T_{c}$, $u\rightarrow 0$ and $\phi\ll 1$. In that
case, we can expand the potential to order $\phi^{4}$ to obtain
\begin{eqnarray}
\label{tc1}
U(\phi)=\frac{1}{2}(1-T)\phi^{2}-\frac{1}{12}\phi^{4}+U_{0}.
\end{eqnarray} 
The wall equation becomes
\begin{eqnarray}
\label{tc2}
\frac{d^{2}\phi}{d\tau^{2}}=-(1-T)\phi+\frac{1}{3}\phi^{3}.
\end{eqnarray}
The uniform solutions are $\phi=0$ and $\phi^{2}=3(1-T)$ yielding
\begin{eqnarray}
\label{tc3}
u=\sqrt{3(1-T)}.
\end{eqnarray}
We can now re-express the potential in the form
\begin{eqnarray}
\label{tc4}
U(\phi)=-\frac{1}{12}(\phi^{2}-u^{2})^{2}.
\end{eqnarray} 
The wall profile is given by
\begin{eqnarray}
\label{tc5}
\int \frac{d\phi}{u^{2}-\phi^{2}}=\frac{\tau}{\sqrt{6}},
\end{eqnarray} 
yielding explicitly
\begin{eqnarray}
\label{tc6}
\phi=u\tanh\left (\frac{u\tau}{\sqrt{6}}\right ).
\end{eqnarray} 
The typical width of the wall, as defined by Eq. (\ref{wp6}), is 
\begin{eqnarray}
\label{tc7}
L=\frac{\sqrt{6}}{u}=\sqrt{2}(1-T)^{-1/2}.
\end{eqnarray}
The width of the wall diverges at the critical point with the
exponent $-1/2$. The wall profile can be rewritten
\begin{eqnarray}
\label{tc8}
\phi=u(T)\tanh\left \lbrack\frac{\tau}{L(T)}\right \rbrack,
\end{eqnarray} 
and the concentration gradient is
\begin{eqnarray}
\label{tc9}
v(\tau)=\frac{u(T)}{L(T)}{\cosh^{-2}\left \lbrack\frac{\tau}{L(T)}\right \rbrack}.
\end{eqnarray} 
The maximum value of the concentration gradient is given by 
\begin{eqnarray}
\label{tc10}
v_{max}=\frac{u^{2}}{\sqrt{6}}=\left (\frac{3}{2}\right )^{1/2}(1-T),
\end{eqnarray} 
and it tends to zero with the exponent $+1$ at the critical temperature. Finally, the surface tension is given by
\begin{eqnarray}
\label{tc11}
\sigma=\left (\frac{2}{3}\right )^{3/2}u^{3}=\lbrack 2(1-T)\rbrack^{3/2},
\end{eqnarray} 
and it vanishes at the critical point with the exponent $+3/2$. These
are the same scalings as in the Cahn-Hilliard theory \cite{ch}.

\subsection{The limit $T\rightarrow 0$}
\label{sec_tz}

Setting $x=u/T$, Eq. (\ref{we10})-b can be rewritten $Tx=\tanh(x)$. For $T\rightarrow 0$, $x\sim 1/T\rightarrow +\infty$ and $u\rightarrow 1$. More precisely, considering the behavior of $\tanh(x)$ for $x\rightarrow +\infty$, we get
\begin{eqnarray}
\label{tz1}
u\simeq 1-2e^{-2/T}.
\end{eqnarray} 
The potential can be rewritten (for $\phi\ge 0$)
\begin{eqnarray}
\label{tz2}
U(\phi)=\phi+Te^{-2\phi/T}-\frac{\phi^{2}}{2}+U_{0}.
\end{eqnarray} 
The typical width of the wall is
\begin{eqnarray}
\label{tz3}
L=2\left (1+\frac{2}{T}e^{-2/T}\right )=2-(1-u)\ln\left (\frac{1-u}{2}\right ),
\end{eqnarray} 
and the maximum value of the concentration gradient is 
\begin{eqnarray}
\label{tz4}
v_{max}=1-(\ln 2) T=1-\frac{2\ln 2}{\ln\left (\frac{1-u}{2}\right )}.
\end{eqnarray} 
Finally, the surface tension is given by (see Appendix \ref{sec_st})
\begin{eqnarray}
\label{tz5}
\sigma=1-{\pi^{2}\over 12}T^{2}=1-\frac{\pi^{2}}{6\ln \left (\frac{1-u}{2}\right )}.
\end{eqnarray} 
For $T=0$, we have $u=1$ and 
\begin{eqnarray}
\label{tz6}
U(\phi)=\phi-\frac{\phi^{2}}{2}-\frac{1}{2}=-\frac{1}{2}(\phi-1)^{2}.
\end{eqnarray} 
The wall profile is solution of
\begin{eqnarray}
\label{tz7}
\frac{d\phi}{d\tau}=1-\phi,
\end{eqnarray} 
leading to
\begin{eqnarray}
\label{tz8}
\phi=1-e^{-\tau}  \qquad (\tau\ge 0).
\end{eqnarray}

\subsection{Simple approximation for $T> 1/2$}
\label{sec_appsup}

If we are relatively close to the critical temperature, we can propose a simple approximation of the wall profile
in the form
\begin{eqnarray}
\label{appsup1}
\phi=u(T)\tanh\left \lbrack\frac{\tau}{L(T)}\right \rbrack,
\end{eqnarray} 
where $u$ and $L$ are given by the {\it exact} expressions 
\begin{eqnarray}
\label{appsup2}
u=\tanh\left (\frac{u}{T}\right ),
\end{eqnarray} 
\begin{eqnarray}
\label{appsup3}
L=\frac{2}{\sqrt{1-\frac{1}{T}(1-u^{2})}}.
\end{eqnarray} 
This Ansatz becomes exact when $T\rightarrow T_{c}$ and it provides a
fair approximation for smaller temperatures (typically $T>1/2$), see
Fig. \ref{phitau7}.

The concentration gradient obtained from Eq. (\ref{appsup1}) is given by
\begin{eqnarray}
\label{appsup4}
v(\tau)=\frac{v_{max}(T)}{\cosh^{2}(\tau/L(T))},
\end{eqnarray}  
with the maximum value 
\begin{eqnarray}
\label{appsup5}
v_{max}=\frac{u}{L}=\frac{u}{2}\sqrt{1-\frac{1}{T}(1-u^{2})}.
\end{eqnarray} 
However, it is more relevant to take for $v_{max}$ the {\it exact} value (\ref{wp8}), i.e.
\begin{equation}
\label{appsup6}
v_{max}=\sqrt{2T\ln\left \lbrack \cosh \left (\frac{u}{T}\right )\right\rbrack-u^{2}}.
\end{equation} 
Typically, Eq. (\ref{appsup4}) with (\ref{appsup5}) gives a better
agreement with the exact solution in the tail of profile while
Eq. (\ref{appsup4}) with (\ref{appsup5}) gives a better agreement in
the core of the profile, see Fig. \ref{tauV7}. Finally, the surface
tension calculated with Eq. (\ref{appsup4}) is given by
$\sigma=\frac{4}{3}v_{max}^2 L$.

\subsection{Simple approximation for $T<1/2$}
\label{sec_appinf}

For sufficiently small temperatures, we can propose a simple
approximation of the wall profile in the form
\begin{eqnarray}
\label{appinf1}
\phi=u(T)\left \lbrack 1-e^{-2\tau/L(u)}\right\rbrack  \qquad (\tau\ge 0),
\end{eqnarray} 
where $u$ and $L$ are given by the {\it exact} expressions
(\ref{appsup2}) and (\ref{appsup3}). The concentration gradient
obtained from Eq. (\ref{appinf1}) is
\begin{eqnarray}
\label{appinf2}
v(\tau)=v_{max}(T){e^{-2|\tau|/L(T)}},
\end{eqnarray}  
with the maximum value $v_{max}=2\frac{u}{L}$. As before, it may be
more relevant to use the exact value (\ref{appsup6}). The surface
tension calculated with Eq. (\ref{appinf2}) is given by
$\sigma=\frac{1}{2}v_{max}^2 L$ but this approximation yields,
unfortunatelly, an asymptotic behaviour for $T\rightarrow 0$ different
from the exact result (\ref{tz5}).

\begin{figure}
\centering
\includegraphics[width=8cm]{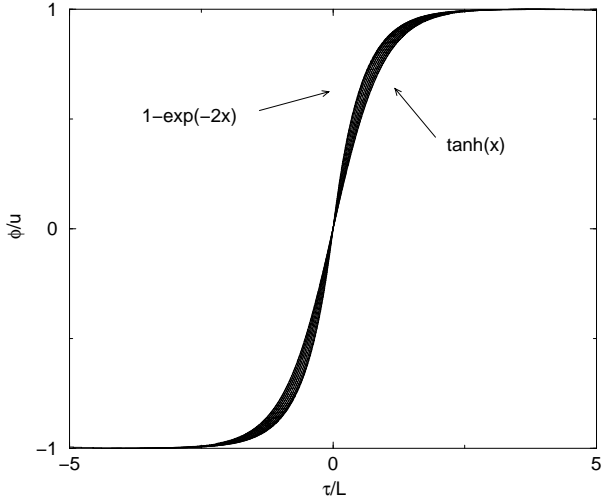}
\caption{Scaled concentration $\phi/u(T)$ as a function of the scaled distance $\tau/L(T)$ for different values of the temperature between $T=0$ and $T=T_{c}=1$. In terms of the
scaled variables, the exact concentration profile is bounded by the
solutions $1-{\rm exp}(-2x)$ for $T=0$ and $\tanh(x)$ for $T=T_{c}$.
}
\label{phinorm}
\end{figure}

\begin{figure}
\centering
\includegraphics[width=8cm]{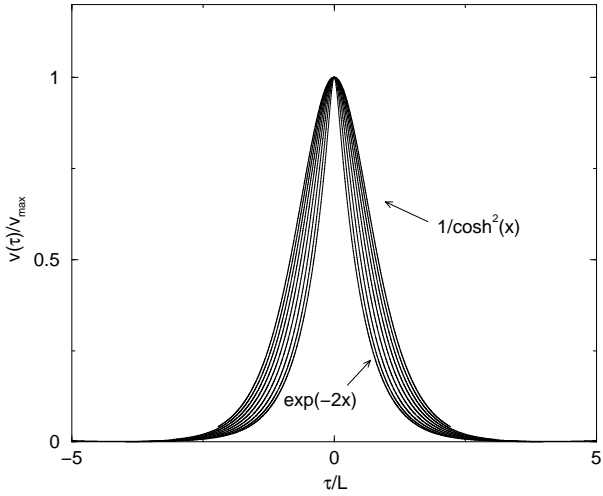}
\caption{Scaled concentration gradient $v/v_{max}(T)$ as a function of the scaled distance $\tau/L(T)$ for different values of the temperature between $T=0$ and $T=T_{c}=1$. In terms of the
scaled variables, the exact profile of concentration gradient is bounded by the
solutions ${\rm exp}(-2x)$ for $T=0$ and $1/\cosh^{2}(x)$ for $T=T_{c}$. }
\label{Vnorm}
\end{figure}

\begin{figure}
\centering
\includegraphics[width=8cm]{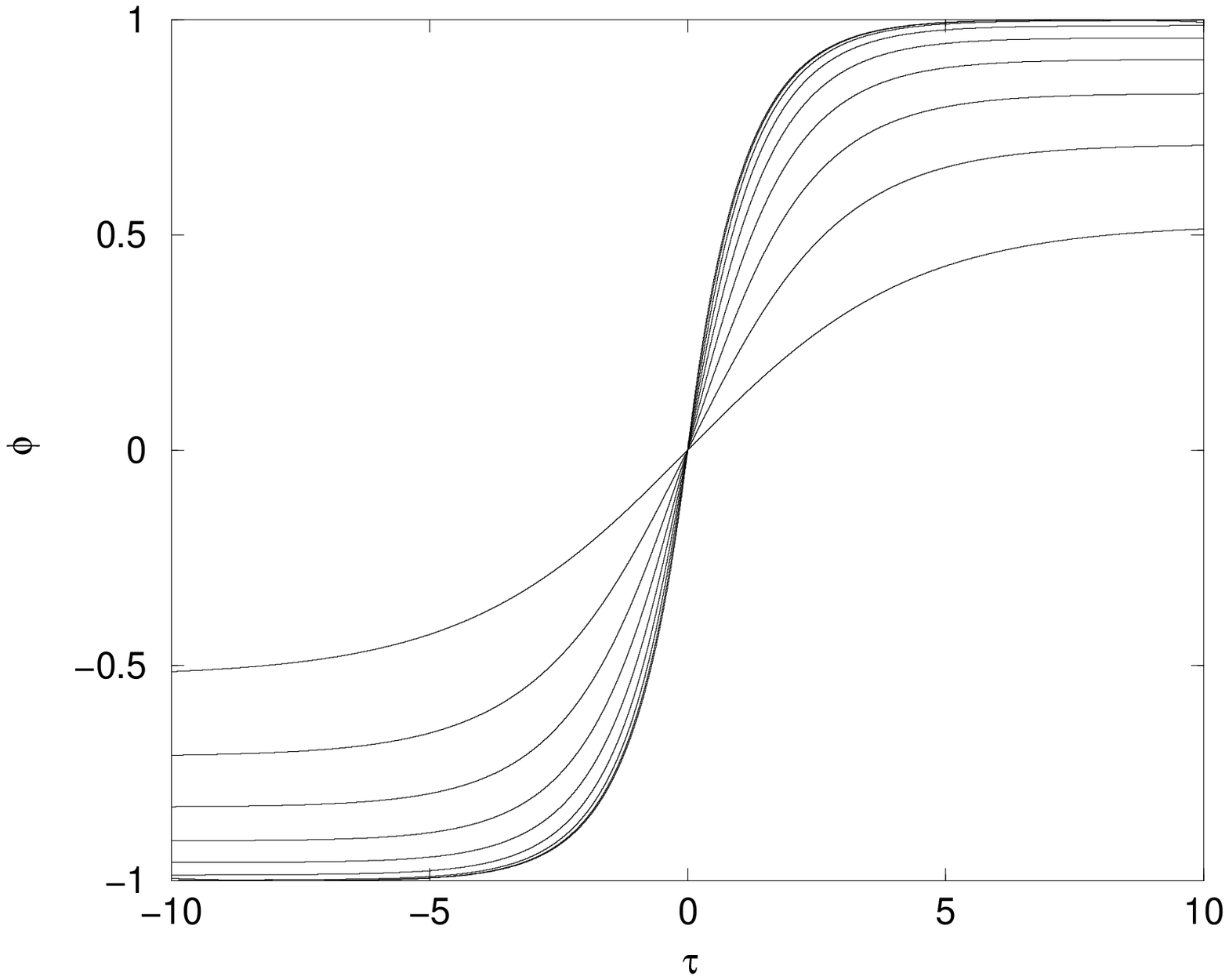}
\caption{Concentration $\phi$ as a function of the distance $\tau$  for different values of the temperature $T=0.1$, $0.2$, $0.3$, $0.4$, $0.5$, $0.6$, $0.7$, $0.8$ and $0.9$. }
\label{allphitau}
\end{figure}

\begin{figure}
\centering
\includegraphics[width=8cm]{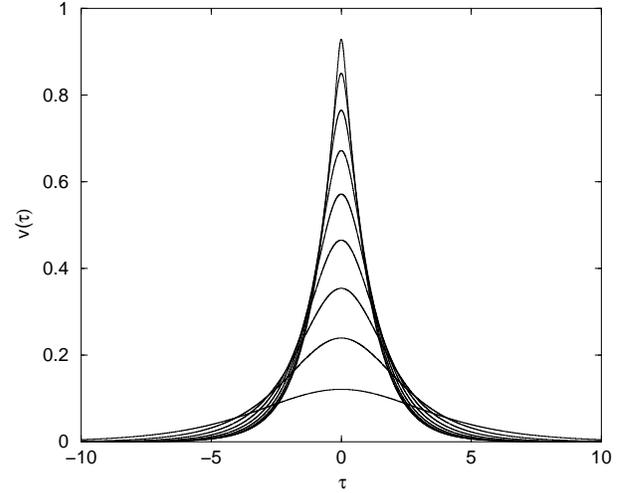}
\caption{Concentration gradient $v$ as a function of the distance $\tau$ for different values of the temperature $T=0.1$, $0.2$, $0.3$, $0.4$, $0.5$, $0.6$, $0.7$, $0.8$ and $0.9$. }
\label{allVtau}
\end{figure}

The Ansatz (\ref{appsup1})-(\ref{appinf1}) have a self-similar
structure as a function of the temperature. Indeed, the functions
$\phi(\tau)/u(T)$ and $v(\tau)/v_{max}(T)$ vs $\tau/L(T)$ have an
invariant profile. The exact solution of the differential equation
(\ref{we8}) has not an exact self-similar structure as shown in
Figs. \ref{phinorm} and \ref{Vnorm} but the region between the
envelopes is relatively thin so that
Eqs. (\ref{appsup1})-(\ref{appinf1}) can be useful approximations for
$T\rightarrow 1$ and $T\rightarrow 0$ respectively. The profiles of
concentration and concentration gradient (in non-scaled variables) are
plotted in Figs.
\ref{allphitau} and  \ref{allVtau}.

\subsection{Match asymptotics}
\label{sec_ma}

For small values of the temperature, one can propose another
approximation of the profiles of concentration and concentration
gradient by using match asymptotics.

\subsubsection{Concentration profile $\phi(\tau)$}
\label{sec_mac}

\begin{figure}
\centering
\includegraphics[width=8cm]{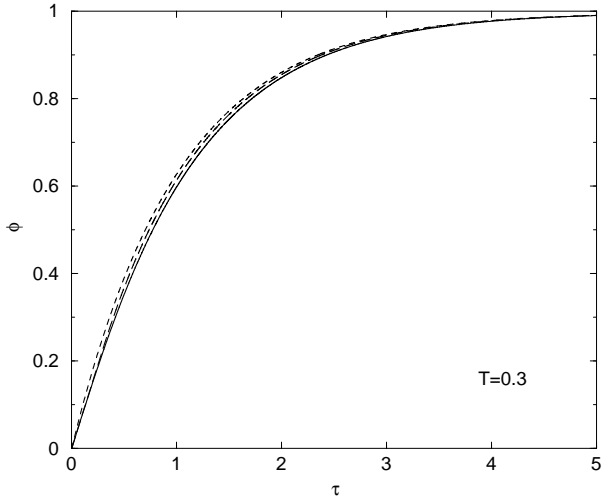}
\caption{Concentration profile for $T=0.3$: exact (solid), approximate self-similar (dotted), match asymptotics (long-dashed). }
\label{phitau3}
\end{figure}

The exact asymptotic behaviors of the concentration profile are given by 
\begin{eqnarray}
\label{mac1}
\phi(\tau)\simeq v_{max}\tau+... \qquad (\tau\rightarrow 0),
\end{eqnarray} 
\begin{eqnarray}
\label{mac2}
\phi(\tau)=u-Ae^{-2\tau/L} \qquad (\tau\rightarrow +\infty),
\end{eqnarray} 
where $u$, $L$ and $v_{max}$ are known functions of the temperature.
We match these two behaviors at a point $x$ where their values and the
values of their first derivative coincide. This yields
\begin{eqnarray}
\label{mac3}
x \, v_{max}=u-Ae^{-2x/L},
\end{eqnarray} 
\begin{eqnarray}
\label{mac4}
v_{max}=\frac{2A}{L}e^{-2x/L}.
\end{eqnarray}
From these relations, we obtain
\begin{eqnarray}
\label{mac5}
x=\frac{u}{v_{max}}-\frac{L}{2},
\end{eqnarray}
\begin{eqnarray}
\label{mac6}
A=\frac{L}{2}v_{max}e^{\frac{2u}{L v_{max}}-1}.
\end{eqnarray}
For $T\rightarrow 0$, we explicitly  find that
\begin{eqnarray}
\label{mac7}
x=(\ln 2) T,
\end{eqnarray}
\begin{eqnarray}
\label{mac8}
A=1-(\ln 2)^2 T^2.
\end{eqnarray}
The exact concentration profile is plotted in Fig. \ref{phitau3} for
$T=0.3$, and compared with the approximate self-similar expression
(\ref{appinf1}) and the expression
(\ref{mac1})-(\ref{mac2}) obtained by match asymptotics.

\subsubsection{Concentration gradient $v(\tau)$}
\label{sec_mav}

\begin{figure}
\centering
\includegraphics[width=8cm]{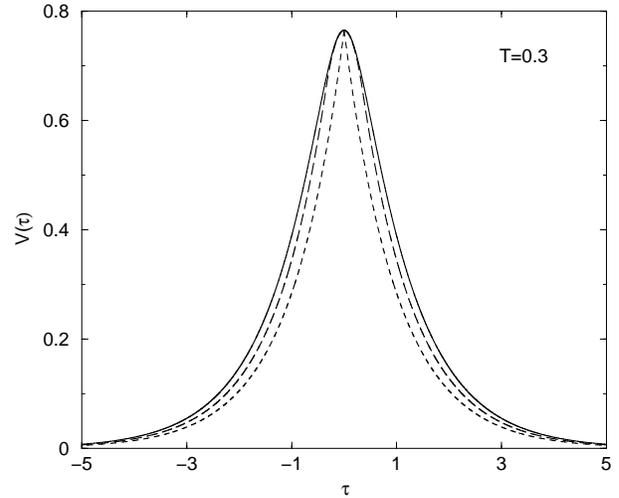}
\caption{Concentration gradient for $T=0.3$: exact (solid), approximate self-similar (dotted), match asymptotics (long-dashed). }
\label{tauV3}
\end{figure}

The exact asymptotic behaviors of the concentration
gradient are given by
\begin{eqnarray}
\label{mav1}
v(\tau)\simeq v_{max}\left\lbrack 1+\frac{1}{2}\left (1-\frac{1}{T}\right )\tau^{2}+...  \right\rbrack  \quad (\tau\rightarrow 0),\nonumber\\
\end{eqnarray} 
\begin{eqnarray}
\label{mav2}
v(\tau)=2\frac{A}{L}e^{-2\tau/L} \qquad (\tau\rightarrow +\infty).
\end{eqnarray} 
We match these two behaviors at a point $x$ where their values and the values of their first derivative coincide. After simplification, we obtain
\begin{eqnarray}
\label{mav3}
x=\frac{-L+\sqrt{L^{2}+\frac{8T}{1-T}}}{2},
\end{eqnarray} 
\begin{eqnarray}
\label{mav4}
A=\frac{L^{2}}{4}v_{max}\left (\frac{1}{T}-1\right )x e^{2x/L}.
\end{eqnarray} 
For $T\rightarrow 0$, we explicitly find that
\begin{eqnarray}
\label{mav5}
x=T,
\end{eqnarray}
\begin{eqnarray}
\label{mav6}
A=1+(1-\ln 2) T.
\end{eqnarray}
The exact concentration gradient is plotted in Fig. \ref{tauV3} for
$T=0.3$, and compared with the approximate self-similar expression
(\ref{appinf2})-(\ref{appsup6}) and the expression (\ref{mav1})-(\ref{mav2}) obtained
by match asymptotics.

\subsection{The curvature radius}
\label{sec_curvature}

To next order in the expansion in $k^{-1}\ll 1$, we must account for
the curvature of the interface. Close to the interface,
$\nabla\psi=\frac{d\psi}{d\xi}{\bf n}$ where ${\bf n}$ is a unit
vector normal to the wall. Introducing the curvature
$r^{-1}=\nabla\cdot {\bf n}$ ($r$ is the curvature radius), we get
$\Delta\psi=d^{2}\psi/d\xi^{2}+r^{-1}d\psi/d\xi$. Therefore, Eq. (\ref{sts5})
becomes at first order
\begin{eqnarray}
\label{curvature1}
\frac{d^{2}\phi}{d\tau^{2}}+\frac{1}{kr}\frac{d\phi}{d\tau}=-U'(\phi)+\chi_{1}/k,
\end{eqnarray} 
where we have written $\chi=\chi_{0}+k^{-1}\chi_{1}+...$ and used the
fact that $\chi_{0}=0$ at leading order (see
Sec. \ref{sec_wp}). Multiplying Eq. (\ref{curvature1}) by
$d\phi/d\tau$ and integrating along the wall we obtain the relation
\begin{eqnarray}
\label{curvature2}
\frac{1}{r}=-\frac{2\chi_{1}u}{\sigma(u)},
\end{eqnarray} 
which shows that the radius of curvature $r$ is constant. Therefore, the
shape of the interface is either a circle (leading to a spot) or a
straight line (leading to a stripe). Equation (\ref{curvature2}) is similar to Laplace's law relating the curvature radius of bubbles to the surface tension and to the difference of pressure between the interface.

\subsection{The free energy}
\label{sec_fe}

The previous results can be recovered by minimizing the `free energy'
(\ref{dyn19})-(\ref{dyn20}). This has been discussed by Bouchet \&
Sommeria \cite{bs} in the context of jovian vortices where the problem
is similar (see Sec. \ref{sec_analogy}). Therefore, their study can be
directly applied to the present situation and we shall only give the
main steps of the analysis.

To leading order in $k^{-1}\rightarrow 0$, the system consists of two
phases with uniform density $\rho_{\pm 1}$ and size $A_{\pm 1}$ (we
call $A=A_{+}+A_{-}$ the total size of the domain). Setting
$\rho=\frac{\sigma_{0}}{2}(1+\phi)$ and
$c=\frac{\lambda\sigma_{0}}{2k^{2}}(1+\phi)$, the free energy $F=E-T_{eff}S+\alpha T_{eff}M$ where $E$ and $S$ are given by Eqs. (\ref{dyn19})-(\ref{dyn20}) and $M$ is the total mass  can be written
\begin{eqnarray}
\label{fe1}
F=A_{1}f(\rho_{1})+(A-A_{1})f(\rho_{-1}),
\end{eqnarray} 
with
\begin{eqnarray}
\label{fe2}
f=\frac{T_{eff}\sigma_{0}}{2}\lbrack -C\phi^{2}+(\alpha-2C)\phi\nonumber\\
+(1+\phi)\ln(1+\phi)+(1-\phi)\ln(1-\phi)\rbrack.
\end{eqnarray} 
The optimal values of  $\rho_{\pm 1}$ and  $A_{\pm 1}$ are obtained by minimizing the free energy (\ref{fe1}). The variations on  $\rho_{\pm 1}$ imply that
\begin{eqnarray}
\label{fe3}
f'(\rho_{\pm 1})=0,\qquad f''(\rho_{\pm 1})>0.
\end{eqnarray} 
and the variations on $A_{1}$ imply that
\begin{eqnarray}
\label{fe4}
f(\rho_{1})=f(\rho_{-1}).
\end{eqnarray} 
This relation expresses the equality of the free energy of the two
phases. The only possibility to satisfy the relations
(\ref{fe3})-(\ref{fe4}) simultaneously is to have $\alpha=2C$ so that
$f(\phi)$ is an odd function. This is equivalent to the solvability
condition (\ref{we7}) leading to $\chi\equiv \alpha/2C-1=0$. Then, it
is straightforward to check that $f'(\rho)=0$ implies that $\phi=\pm
u$ where $u$ is given by
\begin{eqnarray}
\label{fe5}
Cu=\frac{1}{2}\ln\left (\frac{1+u}{1-u}\right )=\tanh^{-1}(u).
\end{eqnarray} 
This returns the relation (\ref{we10})-b. 

To first order in $k^{-1}$, we need to determine the contribution of
the free energy contained in the wall (interface). The free energy per
unit length is given by
\begin{eqnarray}
\label{fe6}
F_{W}=\frac{1}{k}\int_{-\infty}^{+\infty}\left\lbrack h(\rho(\tau))-h(\rho_{\pm 1})\right\rbrack d\tau,
\end{eqnarray} 
where $h(\rho)$ is the density of free energy. Using Eqs. (\ref{dyn19}), (\ref{dyn20}), (\ref{we13}) and (\ref{wp2}), we obtain after simplification
\begin{eqnarray}
\label{fe7}
F_{W}=\frac{\lambda\sigma_{0}^{2}}{4k^{3}}\int_{-\infty}^{+\infty}\left\lbrack {\tilde h}(\phi)-{\tilde h}(\phi_{\pm 1})\right\rbrack d\tau,
\end{eqnarray} 
with
\begin{eqnarray}
\label{fe8}
 {\tilde h}(\phi)=\phi (\tanh(C\phi)-\phi).
\end{eqnarray} 
Using ${\tilde h}(\phi_{\pm 1})={\tilde h}(u)=0$ and
$\tanh(C\phi)\ge\phi$ (see Fig. \ref{tanh}), we find that
$F_{W}>0$. Therefore, minimizing the free energy amounts to minimizing
the length of the interface at fixed area. This gives either a circle
or a straight line and this returns the fact that the curvature radius
is constant (see Sec. \ref{sec_curvature}). This argument also shows
that it is more profitable to form, at equilibrium, a single
``bubble'' of size $A$ rather than several ``droplets'' of smaller
size. However, a configuration with several ``droplets'' can exist as
a non-equilibrium solution of the regularized Keller-Segel model
(\ref{dyn16})-(\ref{dyn17}). These droplets will evolve in time and
merge together to finally form a single ``bubble'' (spot or
stripe). This is similar to a coarsening process in spin systems or to
the aggregation of vortices in 2D decaying turbulence \cite{aggreg}.

\subsection{Bifurcations: spots and stripes}
\label{sec_bif}

\begin{figure}
\centering
\includegraphics[width=8cm]{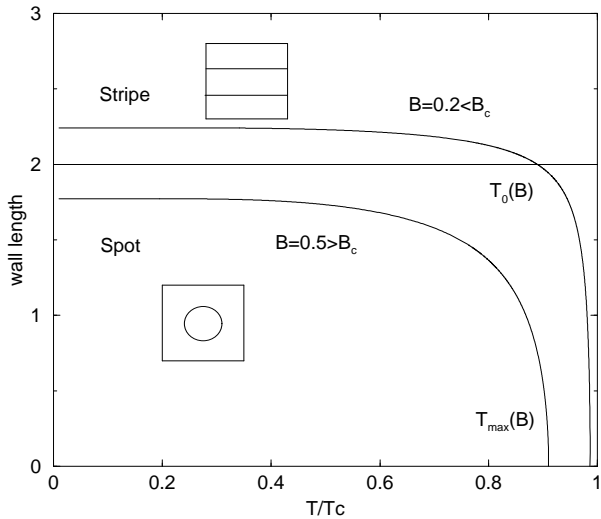}
\caption{Interfacial length of the spot (circle) as a function of the temperature for different values of the parameter $B$ (the area of the domain has been normalized to $A=1$). The spot is selected if its length is smaller than $2$ and the stripe is selected in the other case. If $B>B_{c}=(\pi-2)/\pi$, the spot is always selected. If $B<B_{c}$, the stripe is selected for $T<T_{0}(B)$ and the spot is selected for $T>T_{0}(B)$.}
\label{walllength}
\end{figure}

We shall work in a square domain with total size $A$. We consider periodic
boundary conditions in order to avoid boundary effects. The
equilibrium state consists in two phases with uniform density
$(\rho_{+},A_{+})$ and $(\rho_{-},A_{-})$ with $A_{+}+A_{-}=A$. We introduce the parameter 
\begin{eqnarray}
\label{bif1}
B=1-\frac{2M}{\sigma_{0}A}.
\end{eqnarray} 
Since $0\le M\le \sigma_{0}A$, the parameter $B$ takes values between $-1$ and $+1$. Writing $M=A_{+}\rho_{+}+A_{-}\rho_{-}$ and using Eq. (\ref{we12}), the area of the two phases at equilibrium are given by
\begin{eqnarray}
\label{bif2}
A_{\pm}=\frac{A}{2}\left (1\mp\frac{B}{u}\right ).
\end{eqnarray} 
They are determined by the parameter $B$ (fixed by the total mass) and by the parameter $u$ (fixed by the temperature). Since $0\le A_{\pm}\le A$, we have the inequalities
\begin{eqnarray}
\label{bif3}
 |B|\le u\le 1, \qquad -1\le B\le 1.
\end{eqnarray} 
Since the curvature radius is constant, we have three possible
configurations: (i) a circular domain (spot) of phase $+$ surrounded
by phase $-$: the length of the interface is $2\sqrt{\pi A_{+}}$. (ii)
a circular domain (spot) of phase $-$ surrounded by phase $+$: in that
case, the length of the interface is $2\sqrt{\pi (A-A_{+})}$. (iii) a
stripe of phase $\pm$ and a stripe of phase $\mp$: the length of the
interface is $2\sqrt{A}$. The configuration selected at equilibrium is
the one with the smallest interfacial length (see
Fig. \ref{walllength}).

(i) The spot of phase $+$ surrounded by phase
$-$ will be selected if $A_{+}\le A/\pi$, i.e. $u\le \pi B/(\pi-2)$. This is possible only for $B\ge 0$. In terms of the temperature, this corresponds to $T\ge T_{0}$ where $T_{0}(B)$ is determined by $u_{0}=\tanh(u_{0}/T_{0})$ with $u_{0}=|B|\pi/(\pi-2)$.

(ii) The circular domain of phase $-$ surrounded by phase
$+$ will be selected if $A_{+}\ge (1-1/\pi)A$, i.e. $u\le -\pi B/(\pi-2)$. This is possible only for $B\le 0$. In terms of the temperature, this corresponds to $T\ge T_{0}$.

(iii) The stripes will be selected for $T<T_{0}$. This is possible
only for $|B|\le B_{c}\equiv (\pi-2)/\pi$ (i.e. $u_{0}\le 1$). For $B>0$, the stripe $+$ has the smallest area ($A_{+}\le A_{-}$) and this is the opposite for $B<0$.

Finally, we note that the condition $|B|\le u$ implies that $T\le T_{max}(B)\le T_c$ where $T_{max}(B)$ is determined by $|B|=\tanh(|B|/T_{max})$.

\begin{figure}
\centering
\includegraphics[width=8cm]{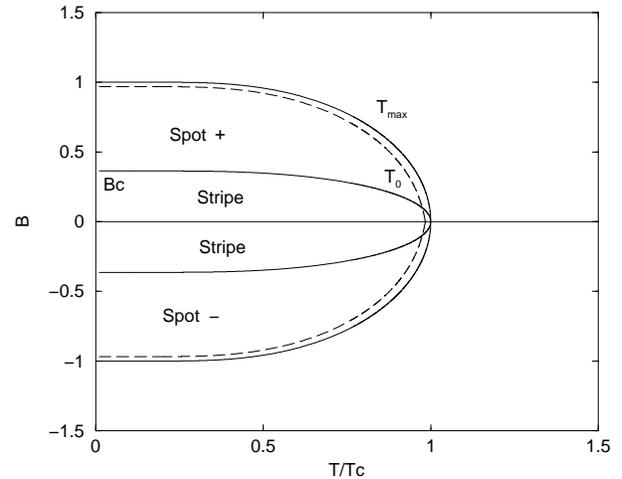}
\caption{Phase diagram of the regularized chemotactic model (\ref{dyn16})-(\ref{dyn17}) showing the bifurcation between `spots' and `stripes' as a function of the control parameters $(B,T)$. In a bounded domain, the solutions exist only for $T\le T_{max}(B)$. We have drown the line of transition $T_{0}(B)$ between the two structures. Finally, the dashed line corresponds to the domain of validity of our perturbative expansion; it has been plotted for $Ak^{2}=1000$. }
\label{diagphase}
\end{figure}

The phase diagram is represented in Fig. \ref{diagphase}. This is the
counterpart to the diagram obtained by Bouchet \& Sommeria \cite{bs} for
jovian vortices. The `spots' are the equivalent of the `vortices' and
the `stripes' are the equivalent of the `jets'. The main difference
(beyond the context and the interpretation of the solutions) is that
the control parameter in our case is the effective temperature $T$
(``canonical'' situation) while their control parameter is the energy
$E$ (microcanonical situation).
 
Finally, concerning the domain of validity, our approach assumes that
the domain size is larger than the interfacial width $L(u)/k$ given by
(\ref{wp6}) so that $A_{\pm}\ge (2L/k)^{2}$. This is satisfied on the
left of the dashed line in Fig. \ref{diagphase} corresponding to
\begin{eqnarray}
\label{bif4}
|B|\le u\left (1-\frac{8L(u)^{2}}{A k^{2}}\right ).
\end{eqnarray}

\section{Analogy with Jupiter's great red spot}
\label{sec_analogy}

\subsection{The physical context}
\label{sec_pc}

Jupiter's Great Red Spot (GRS) is probably the most famous example of
vortex structures found in planetary atmospheres. The presence of this
spot was first reported by Robert Hooke in 1664 in the first issue of
the Philosophical Transaction of the Royal Society. The GRS is a large
``eye'' that dominates the southern hemisphere of the planet. It is an
oval-shaped anticyclone with size 26000km by 13000km. Its breadth is
about one hundred times its height so it can be considered
approximately two-dimensional. It resides in a zonal shear at latitude
23S where the velocity changes sign, is elongated along the shear zone
and is of the same sign as the background shear (these are relatively
general rules observed for other vortices). It stays at the same
latitude but slowly drifts in longitude.  Morphologically, the GRS has
a striking annular structure with a quiet center surrounded by a thin
intense jet.  The concentration of winds in an annulus is consistent
with the fact that the GRS is much larger than the atmosphere's radius
of deformation (the jet's width scales with the Rossby radius of
deformation).

Some authors have tried to describe this spot as a soliton solution of
a modified KdV equation but this interpretation as a weakly nonlinear
structure faces several drawbacks: the jet structure is not reproduced
and the soliton theory predicts an interpenetration without change of
structure while the main interaction between vortices is a merging
process. In contrast, Marcus \cite{marcus} argues that a permanent
vortex can coexist with turbulence and that most of the properties of
the Jovian vortices can be easily explained and understood with the
quasi-geostrophic (QG) theory.  In addition, he emphasizes the jet
structure of the GRS and shows that an annular jet is the natural
structure of a vortex with a uniform potential vorticity inside and
outside the spot. A physical justification of this construction was
given by Sommeria et al. \cite{nore} in terms of statistical
mechanics. A statistical theory of 2D turbulence has been developed by
Miller \cite{miller} and Robert \& Sommeria \cite{rs} following and
generalizing the pioneering work of Onsager \cite{onsager} on point
vortices. This theory is able to account for the typical vortices
(monopoles, dipoles, tripoles,...) observed in two-dimensional flows
\cite{class}. In the QG model, and in the limit of
small Rossby radius, PV mixing (entropic effects) with constraints on
the energy leads to an equilibrium state that consists of two phases
with uniform PV in contact separated by a strong jet. This precisely
account for the morphology of the GRS. This model has been further
developed by Bouchet \& Sommeria \cite{bs} with quantitiative
applications to Jovian vortices. Another model has been proposed by
Turkington et al. \cite{turkington}. It predicts the emergence of a
vortex solution at the correct latitude but it does not reproduce the
annular jet structure of the GRS \cite{physicaD}.

\subsection{Statistical mechanics of the quasi-geostrophic equations}
\label{sec_qg}

The quasi-geostrophic equations appropriate to the dynamics
of geophysical flows \cite{pedlosky} can be written:
\begin{eqnarray}
\label{qg1}
\frac{\partial q}{\partial t}+{\bf u}\cdot \nabla q=0,
\end{eqnarray} 
\begin{eqnarray}
\label{qg2}
q=-\Delta\psi+\frac{\psi}{R^{2}}-Rh(y), \quad {\bf u}=-\hat{\bf z}\times \nabla\psi.
\end{eqnarray} 
Here, $q$ is the potential vorticity (PV) and $\psi$ the stream
function ($\hat{\bf z}$ is a unit vector normal to the two-dimensional
flow). We have assumed that the topography $Rh(y)$ scales with the
Rossby radius $R$.  The Q.G. equations admit an infinite number of
stationary solutions specified by any relationship $q=f(\psi)$. For
given initial conditions, the statistical theory selects the most
probable state consistent with the constraints imposed by the
dynamics. It is obtained by maximizing a mixing entropy at fixed
energy and Casimir constraints \cite{miller,rs,csr}. 

Let us consider the situation where the fine-grained PV $q$ takes only
two values $\lbrace a_{-1},a_{1}\rbrace$. It is convenient to set
$(a_{1}-a_{-1})/2=1$ and $(a_{1}+a_{-1})/2=B$. We also choose the
Gauge condition on $\psi$ such that $\langle q\rangle=0$
\cite{bs}. Then, the total area occupied by level $a_{1}$ is ${\cal
A}=(1-B)/2$ (the total area of the domain is unity). In this
two-levels case \cite{miller,rs,csr}, the mixing entropy is given by
\begin{eqnarray}
\label{qg3}
S=-\int \lbrack p\ln p+(1-p)\ln(1-p)\rbrack d{\bf r},
\end{eqnarray} 
where $p({\bf r})$ is the local distribution of level $a_{1}$.  The coarse-grained PV is $\overline{q}=p a_{1}+(1-p) a_{-1}$. The extremization of (\ref{qg3}) at fixed energy
\begin{eqnarray}
\label{qg4}
E=\frac{1}{2}\int (\overline{q}+h)\psi d{\bf r}=\frac{1}{2}\int \left\lbrack (\nabla\psi)^{2}+\frac{\psi^{2}}{R^{2}}\right \rbrack d{\bf r},
\end{eqnarray} 
and total area ${\cal A}=\int p({\bf r})d{\bf r}$ leads to a $q-\psi$ relation of the form
\begin{eqnarray}
\label{qg5}
\overline{q}=B-\tanh\left (\alpha-\frac{C\psi}{R^{2}}\right ),
\end{eqnarray} 
where $\alpha$ and $C$ are Lagrange multipliers introduced in the variational problem $\delta S-2\alpha\delta {\cal A}+\frac{C}{R^{2}}\delta E=0$.

\begin{figure}
\centering
\includegraphics[width=8cm]{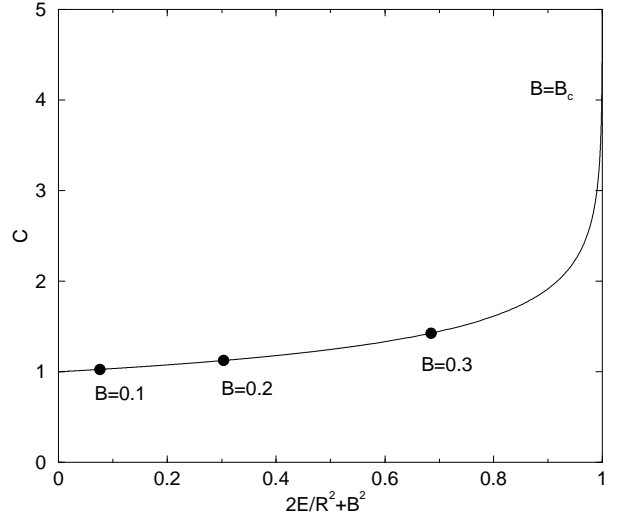}
\caption{Caloric curve giving the inverse temperature $C$ as a function of the energy $E$ for jovian vortices in the limit of small radius of deformation $R\rightarrow 0$. This curve is obtained from Eq. (\ref{jet9})-(\ref{we10}) in the absence of topography. In terms of the variable $2E/R^{2}+B^{2}$, this curve is independent on $B$. We have indicated by a `bullet' the point of bifurcation 
(corresponding to $u=\pi B/(\pi-2)$) between a vortex (spot, left) and a straight jet (stripe, left), for different values of $B$: $0.1$,
$0.2$ and $0.3$.  }
\label{caloric}
\end{figure}

Robert \& Sommeria \cite{rsmepp} have proposed a parameterization of
2D turbulence in the form of a relaxation equation for the
coarse-grained PV $\overline{q}({\bf r},t)$. This parameterization is
based on a Maximum Entropy Production Principle (MEPP). The diffusion current (due to turbulent mixing) is assumed to maximize
the rate of entropy production $\dot S$ while conserving all the
constraints imposed by the dynamics. In the two-levels case, this yields a system of equations of the form
\begin{eqnarray}
\label{qg6}
\frac{\partial \overline{q}}{\partial t}+{\bf u}\cdot \nabla \overline{q}=\qquad\qquad\qquad\qquad\qquad\qquad\nonumber\\
\nabla\cdot \left\lbrack D\left (\nabla \overline{q}+\beta(t)(a_{1}-\overline{q})(\overline{q}-a_{-1})\nabla\psi\right )\right\rbrack,
\end{eqnarray} 
\begin{eqnarray}
\label{qg7}
\beta(t)=-\frac{\int D\nabla\overline{q}\cdot\nabla\psi \, d{\bf r}}{\int D(a_{1}-\overline{q})(\overline{q}-a_{-1})(\nabla\psi)^{2}\, d{\bf r}},
\end{eqnarray} 
\begin{eqnarray}
\label{qg2b}
q=-\Delta\psi+\frac{\psi}{R^{2}}-Rh(y).
\end{eqnarray} 
Interestingly, these drift-diffusion equations are similar to the
regularized chemotactic model (\ref{dyn16})-(\ref{dyn17}). In this
analogy, the coarse-grained PV $\overline{q}$ plays the role of the
bacterial concentration and the stream function $\psi$ the role of the
chemical $c$. This analogy was noted in \cite{crrs}. An important
difference, however, is that in 2D turbulence the energy is conserved
so that the inverse temperature $\beta(t)$ evolves in time. By contrast, in
the chemotactic problem, we are in a situation where the
`effective temperature' $T_{eff}=1/\beta$ is fixed. Therefore, 2D
turbulence corresponds to a microcanonical situation (where we
maximize the entropy $S$ at fixed energy $E$) while chemotaxis
corresponds to a canonical situation (where we minimize an effective
free energy $F=E-T_{eff}S$). These two situations are described in
\cite{gen}. For long-range interactions ($k=0$ or $R\rightarrow
+\infty$), the `ensembles' are generically inequivalent. However, in
the limit that we consider here, corresponding to a short-range
interaction, they become equivalent, i.e. the $\beta(E)$ curve is
univalued (see Fig. \ref{caloric}).

The equilibrium problem (\ref{qg5})-(\ref{qg2b}) has been studied in
the limit of small Rossby radius $R\rightarrow 0$. The original idea
dates back to Sommeria et al. \cite{nore} who understood that, in this
limit, the solution is made of two uniform PV regions separated by a
strong jet. This model has been developed quantitatively by Bouchet \&
Sommeria \cite{bs} with comparison to jovian data. Our study of the
regularized chemotactic problem has been directly insiped by these
works. In complement, we have provided in Sec. \ref{sec_dwall} a more
thorough study of the wall equation with useful asymptotic expansions
and analytical approximations. Due to the analogy between the two
problems, these results can also be relevant to describe the jet
structure of jovian vortices, like GRS. In this respect, we recall
that the GRS corresponds to a typical parameter $u$ in the range
$0.92\le u\le 1$ \cite{bs} so that the limit $u\rightarrow 1$ or
$T\rightarrow 0$ of our study (Secs. \ref{sec_tz}, \ref{sec_appinf}
and \ref{sec_ma}) is particularly interesting in that respect. To
strenghten the comparison between the two problems (chemotaxis and
jovian vortices), we briefly recall the main lines of the study of
Bouchet \& Sommeria \cite{bs} and provide some complementary
discussion.

\subsection{Domain wall theory of Jupiter's great red spot}
\label{sec_wgrs}

\subsubsection{The jet equation}
\label{sec_jet}

Combining Eqs. (\ref{qg2b}) and (\ref{qg5}), we find that the
streamfunction satisfies the meanfield equation
\begin{eqnarray}
\label{jet1}
-\Delta\psi+\frac{\psi}{R^{2}}-Rh(y)=B-\tanh\left (\alpha-\frac{C\psi}{R^{2}}\right ).
\end{eqnarray} 
We shall solve this equation perturbatively as an expansion in powers
of $R$. We only give the main lines and refer to Bouchet \& Sommeria
\cite{bs} for more details and developements. To first order, we obtain the jet equation
\begin{eqnarray}
\label{jet2}
-\frac{d^{2}\psi}{d\xi^{2}}-\frac{1}{r}\frac{d\psi}{d\xi}+\frac{\psi}{R^{2}}-Rh(y)=B-\tanh\left
(\alpha-\frac{C\psi}{R^{2}}\right ),\nonumber\\
\end{eqnarray} 
where $r$ is the curvature radius.  Introducing the notations
\begin{eqnarray}
\label{jet3}
\tau=\xi/R,\qquad \phi=\frac{\psi}{R^{2}}-\frac{\alpha}{C},
\end{eqnarray} 
we get
\begin{eqnarray}
\label{jet4}
\frac{d^{2}\phi}{d\tau^{2}}+\frac{R}{r}\frac{d\phi}{d\tau}+Rh(y)=-\tanh(C\phi)+\phi+\frac{\alpha}{C}-B.\nonumber\\
\end{eqnarray} 
To leading order in $R\ll 1$, the foregoing equation reduces to 
\begin{eqnarray}
\label{jet5}
\frac{d^{2}\phi}{d\tau^{2}}=-\tanh(C\phi)+\phi+\frac{\alpha_{0}}{C}-B.
\end{eqnarray}
The condition of solvability implies that $\alpha_{0}=CB$ so that the jet equation takes the form 
\begin{eqnarray}
\label{jet6}
\frac{d^{2}\phi}{d\tau^{2}}=-\tanh(C\phi)+\phi=-U'(\phi).
\end{eqnarray}
This is the same equation as Eq. (\ref{we8}) with $C=1/T$. In the
present context, $\phi$ is related to the streamfunction and
$v=d\phi/d\tau$ to the jet velocity. Therefore, the figures
representing the profile of concentration gradient in
Sec. \ref{sec_dwall} give the jet velocity profile in the present
context. The PV and streamfunction in the two phases are
\begin{eqnarray}
\label{jet7}
q_{\pm}=\frac{\psi_{\pm}}{R^{2}}=B\pm u.
\end{eqnarray}
Their area is given by
\begin{eqnarray}
\label{jet8}
A_{\pm}=\frac{1}{2}\left (1\mp\frac{B}{u}\right ),
\end{eqnarray}
which is similar to Eq. (\ref{bif2}).  Finally, in the present
context, the parameter $u$ is determined by the energy according to
the relation
\begin{eqnarray}
\label{jet9}
E=\frac{1}{2}R^{2}(u^{2}-B^{2}).
\end{eqnarray}

\subsubsection{The underlying shear}
\label{sec_shear}

To first order in $R$, the jet equation can be written
\begin{eqnarray}
\label{shear1}
\frac{d^{2}\phi}{d\tau^{2}}+\frac{R}{r}\frac{d\phi}{d\tau}+Rh(y)=-U'(\phi)+R\alpha_{1},
\end{eqnarray} 
where we have written $\alpha=\alpha_{0}+R\alpha_{1}+...$ and used
$\alpha_{0}=CB$.  Far from the jet, we have
\begin{eqnarray}
\label{shear2}
R h(y)=-U'(\phi)+R\alpha_{1}.
\end{eqnarray} 
Writing $\phi=u+R\delta\phi$, we get $\delta\phi=(\alpha_{1}-h(y))/U''(u)$ so that the velocity ${\bf v}_{shear}=\delta \phi'(y) {\bf e}_{x}$ is given by 
\begin{eqnarray}
\label{shear3}
v_{shear}=\frac{h'(y)}{-U''(u)}=\frac{h'(y)}{1-C(1-u^{2})}.
\end{eqnarray} 
Now, the analysis of the jet equation in Sec. \ref{sec_wp} shows that
the function appearing in the denominator of Eq. (\ref{shear3}) is
related to the jet width (\ref{wp6}) so that we can write
\begin{eqnarray}
\label{shear4}
v_{shear}(y)=\frac{1}{4} h'(y)L(u)^{2}.
\end{eqnarray}

\subsubsection{The curvature-topography relation}
\label{sec_ct}

\begin{figure}
\centering
\includegraphics[width=8cm]{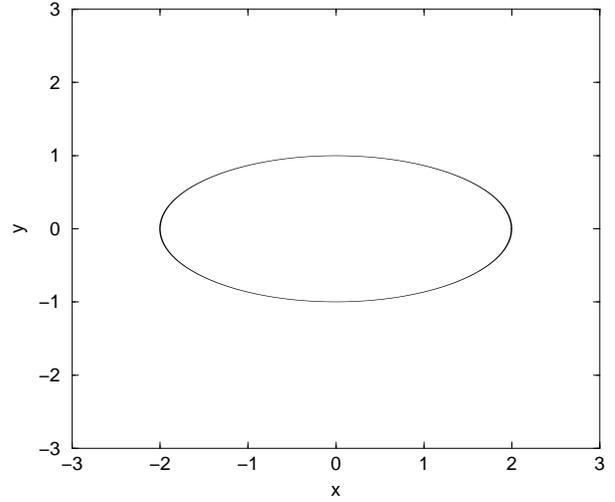}
\caption{Elliptic vortex with aspect ratio of $2$ above a topography of the form (\ref{ct2}).  }
\label{ellipse}
\end{figure}

Finally, multiplying Eq. (\ref{shear1}) by $d\phi/d\tau$ and integrating accross the jet, we obtain
\begin{eqnarray}
\label{ct1}
\frac{e(u)}{r}=u(h(y)-\alpha_{1}),
\end{eqnarray} 
which relates the radius of curvature $r$ to the underlying topography
$h(y)$. For a given topography, Eq. (\ref{ct1}) determines the form
of the jet. This problem has been studied in detail in Bouchet \&
Sommeria \cite{bs} in the case of a quadratic topography. As another
example, we consider here the inverse problem: given the form of
the jet, find the corresponding topography. We consider the case of
an elliptical vortex (see Fig. \ref{ellipse}) because this is a relatively good representation
of Jupiter's great red spot and we can obtain analytical results in
that case.

The topography leading to an elliptic vortex has the form
\begin{eqnarray}
\label{ct2}
h(y)=\frac{H}{\left\lbrack 1+\left ({y}/{L}\right)^{2}\right\rbrack^{3/2}},
\end{eqnarray} 
where $L$ and $H$ are typical horizontal and vertical
lenght scales. Assuming that this relation holds for $|y|\rightarrow
+\infty$, we must take $\alpha_{1}=0$ in Eq. (\ref{ct1}) to have a
vanishing curvature at infinity where $h\rightarrow 0$. Therefore, the curvature radius of the vortex
is given by
\begin{eqnarray}
\label{ct3}
\frac{L\chi}{r}=\left\lbrack 1+\left ({y}/{L}\right)^{2}\right\rbrack^{-3/2},
\end{eqnarray} 
where we have defined $\chi=e(u)/uHL$. This is the equation of an ellipse with major and minor semi-axis: 
\begin{eqnarray}
\label{ct4}
a=\frac{L\chi}{1-\chi^{2}}, \quad b=\frac{L\chi}{\sqrt{1-\chi^{2}}}.
\end{eqnarray} 
These relations assume that $\chi\le 1$. Now, in the case of GRS, the aspect ratio 
\begin{eqnarray}
\label{ct5}
\frac{a}{b}=\frac{1}{\sqrt{1-\chi^{2}}},
\end{eqnarray}
is close to $2$ leading to $\chi=\sqrt{3}/2$. The  major and minor semi-axis
are then given by $a=2\sqrt{3}L$ and $b=\sqrt{3}L$. 

Consider now the limit $\chi\rightarrow 0$. In that case, the width of the vortex is small with respect to the horizontal topographic length ($b\ll L$) and we can make the quadratic approximation  
\begin{eqnarray}
\label{ct6}
h(y)\simeq H-\frac{3}{2}\frac{H}{L^{2}}y^{2}.
\end{eqnarray} 
This is similar to the situation considered by Bouchet \& Sommeria \cite{bs}. Their parameter $d$ is related to our parameter $\chi$ by $d=(3/2)\chi^{2}$. For a quadratic topography with  $(\chi,d)\rightarrow 0$, our study shows that the vortex shape is an ellipse whose  major and minor semi-axis are given by Eqs. (\ref{ct4}). In particular, the aspect ratio behaves like 
\begin{eqnarray}
\label{ct7}
\frac{a}{b}=1+\frac{d}{3}+...\qquad (d\rightarrow 0).
\end{eqnarray} 
This formula is not applicable for a vortex with aspect ratio $\sim 2$
like GRS. Bouchet \& Sommeria \cite{bs} assume a quadratic topography
and solve the curvature-topography relation numerically. For large
values of $d$, close to its maximal value $d_{max}=4/9$, the vortex is
not an ellipse. Alternatively, if we assume a topography of the form
(\ref{ct2}) we find an elliptic vortex for all the values of
$\chi\le 1$. Therefore, the form of the vortex is relatively sensitive to
the underlying topography.

\section{Conclusion}
\label{sec_mf}

In this paper, we have studied the equilibrium states of a regularized
version of the Keller-Segel model describing the chemotactic
aggregation of bacterial colonies. The regularization is justified
physically in order to avoid the formation of singularities (Dirac
peaks) during the dynamics and obtain smooth density profiles (clumps)
instead. This regularization accounts for finite size effects and
close packing effects. In that case, an equilibrium state exists for
any value of the control parameter (this is similar to considering a
gas of self-gravitating fermions in astrophysics to avoid complete
gravitational collapse \cite{fermions}). We have studied furthermore a
limit of high degradation $k\rightarrow +\infty$. In previous works
\cite{crs,sc,lang,post,tcoll,banach,sopik,virial}, the opposite limit
$k=0$ (no degradation) was considered instead. The intermediate case of
a finite degradation must be studied numerically (in
preparation). However, the asymptotic limit $k\rightarrow +\infty$
allows us to obtain analytical results that permit to have a clear
picture of the bifurcation diagram (between spots and stripes) in
parameter space. Furthermore, our approach is exact in one dimension,
for any value of the degradation rate $k$. Our study shows that the
physics of the problem is sensibly different whether $k=0$ or $k\neq
0$.

We have also discussed the analogy between the organization of
bacteria (in stripes and spots) in the chemotactic problem and the
organization of two-dimensional turbulent flows (in jets and vortices)
in the jovian atmosphere. These apparently completely different systems are
however described by relatively similar equations so that an
interesting analogy can be developed between the two. In this analogy,
the jet structure of Jupiter's great red spot can be seen as a `domain
wall' that is similar to the interface separating two phases in
contact, as in the biological problem.

\appendix

\section{Stability analysis}
\label{sec_stab}

\begin{figure}
\centering
\includegraphics[width=8cm]{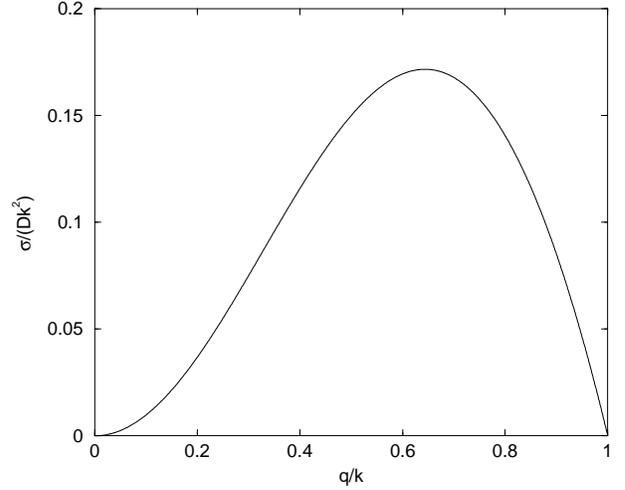}
\caption{Growth rate of the perturbation as a function of the wavenumber for $T=1/2<T_{c}$.}
\label{growthrate}
\end{figure}

We study the linear dynamical stability of an infinite and homogeneous
solution of the regularized Keller-Segel model
(\ref{dyn16})-(\ref{dyn17}). The unperturbed solution is such that
\begin{eqnarray}
\label{stab1}
k^{2}c=\lambda\rho.
\end{eqnarray} 
Linearizing Eqs. (\ref{dyn16})-(\ref{dyn17}) around this steady state and writing the perturbation as $\delta\rho=\delta\hat{\rho}e^{i{\bf q}\cdot {\bf r}}e^{\sigma t}$, $\delta c=\delta\hat{c}e^{i{\bf q}\cdot {\bf r}}e^{\sigma t}$, we obtain
\begin{eqnarray}
\label{stab2}
\chi\rho(1-\rho/\sigma_{0})q^{2}\delta\hat{c}-(Dq^{2}+\sigma)\delta\hat{\rho}=0,
\end{eqnarray} 
\begin{eqnarray}
\label{stab3}
(q^{2}+k^{2})\delta\hat{c}-\lambda\delta\hat{\rho}=0. 
\end{eqnarray} 
This system of equations admits non-trivial solutions only if the determinant is zero yielding the dispersion relation
\begin{eqnarray}
\label{stab4}
\sigma=q^{2}\left (\frac{\chi\lambda\rho (1-\rho/\sigma_{0})}{q^{2}+k^{2}}-D\right ).
\end{eqnarray} 
The system is unstable if $\sigma>0$ and stable otherwise.  A
necessary condition of instability is
\begin{eqnarray}
\label{stab5}
\frac{\chi}{D}\lambda\rho (1-\rho/\sigma_{0})-k^{2}\ge 0.
\end{eqnarray} 
If this condition is fulfilled, the unstable wavenumbers are
\begin{eqnarray}
\label{stab6}
q^{2}\le \frac{\chi}{D}\lambda\rho (1-\rho\sigma_{0})-k^{2}\equiv q_{max}^{2}.
\end{eqnarray} 
For $u=0$, i.e. $\rho=\sigma_0/2$, we find that the instability
criterion (\ref{stab5}) corresponds to
\begin{eqnarray}
\label{stab7}
T\le T_{c}=1.
\end{eqnarray}
Therefore, the uniform phase $u=0$ is stable for $T>T_{c}$ and
unstable for $T<T_{c}$ where it is replaced by a `stripe' or a `spot'
(see Fig. \ref{utnew}). The unstable wavenumbers are 
\begin{eqnarray}
\label{stab8}
q^{2}\le (C-1)k^{2}\equiv q_{max}^{2},
\end{eqnarray} 
where we recall that $C=1/T$.  The growth rate (see
Fig. \ref{growthrate}) can be written
\begin{eqnarray}
\label{stab9}
\sigma=Dq^{2}\left (\frac{Ck^{2}}{q^{2}+k^{2}}-1\right ).
\end{eqnarray} 
The maximum growth rate is obtained for
\begin{eqnarray}
\label{stab10}
q_{*}^{2}=k^{2}(\sqrt{C}-1),
\end{eqnarray} 
and its value is 
\begin{eqnarray}
\label{stab11}
\sigma_{*}=Dk^{2}(\sqrt{C}-1)^{2}.
\end{eqnarray}

\section{Surface tension for $T\rightarrow 0$}
\label{sec_st}

Using Eqs. (\ref{wp9}) and (\ref{wp1}), the surface tension can be
written
\begin{eqnarray}
\label{st1}
\sigma=2\int_{0}^{u}(\phi^{2}-2T\ln\left\lbrack\cosh (\phi/T)\right\rbrack\qquad\qquad\nonumber\\
-u^{2}+2T\ln\left\lbrack\cosh (u/T)\right\rbrack)^{1/2}d\phi.
\end{eqnarray} 
Using 
\begin{eqnarray}
\label{st2}
\ln\left\lbrack\cosh (\phi/T)\right\rbrack=\phi/T-\ln 2+\ln(1+e^{-2\phi/T}),
\end{eqnarray} 
and considering the limit $T\rightarrow 0$, we obtain
\begin{eqnarray}
\label{st3}
\sigma=2\int_{0}^{1}\sqrt{(1-\phi)^{2}-2T\ln (1-e^{-2/T}+e^{-2\phi/T})}d\phi. \nonumber\\
\end{eqnarray} 
Setting $x=1-\phi$, this can be rewritten
\begin{eqnarray}
\label{st4}
\sigma=2\int_{0}^{1} x\sqrt{1-{2T\over x^{2}}\ln \left \lbrack 1+e^{-2/T}(e^{2x/T}-1)\right\rbrack}dx. \nonumber\\
\end{eqnarray} 
For $T\rightarrow 0$, we obtain
\begin{eqnarray}
\label{st5}
\sigma=2\int_{0}^{1} x \left (1-{T\over x^{2}}\ln \left \lbrack 1+e^{-2/T}(e^{2x/T}-1)\right\rbrack \right )dx. \nonumber\\
\end{eqnarray} 
Setting $y=2x/T$, we find that
\begin{eqnarray}
\label{st6}
\sigma=1-2T\int_{0}^{2/T} \ln \left \lbrack 1+e^{-2/T}(e^{y}-1)\right\rbrack {dy\over y}. \nonumber\\
\end{eqnarray} 
Setting $x=2/T-y$, we get
\begin{eqnarray}
\label{st7}
\sigma=1-2T\int_{0}^{2/T} \ln \left \lbrack 1+e^{-2/T}(e^{2/T-x}-1)\right\rbrack {dx\over 2/T-x}. \nonumber\\
\end{eqnarray} 
For $T\rightarrow 0$, we finally obtain
\begin{eqnarray}
\label{st8}
\sigma=1-T^{2}\int_{0}^{+\infty} \ln \left ( 1+e^{-x}\right ) dx. \nonumber\\
\end{eqnarray} 
Using 
\begin{eqnarray}
\label{st9}
\int_{0}^{+\infty} \ln \left ( 1+e^{-x}\right ) dx={\pi^{2}\over 12},
\end{eqnarray} 
we establish Eq. (\ref{tz5}).

\end{document}